\documentclass{article}

\usepackage[english]{babel}
\usepackage{times}
\usepackage[T1]{fontenc}
\usepackage{comment}
\usepackage{natbib}
\usepackage{pifont}
\usepackage{tikz}
\usetikzlibrary{shapes,fit, bayesnet, graphs, arrows, calc, positioning, matrix}

\usepackage{amsthm}
\usepackage{amsmath,amssymb,enumerate,rotate,multirow}
\usepackage{bm,latexsym,mathrsfs,bbm,amsfonts}
\usepackage{color}
\usepackage{verbatim}
\usepackage{subfig}
\usepackage{alphalph}

\usepackage{morefloats}
\usepackage{makecell}
\usepackage{pifont}
\usepackage{diagbox}
\usepackage{authblk}

\usepackage{xr}
\makeatletter

\newcommand*{\addFileDependency}[1]{
\typeout{(#1)}
\@addtofilelist{#1}
\IfFileExists{#1}{}{\typeout{No file #1.}}
}\makeatother

\newcommand*{\myexternaldocument}[1]{%
\externaldocument{#1}%
\addFileDependency{#1.tex}%
\addFileDependency{#1.aux}%
}

\myexternaldocument{Longitudinal_IRT_SM}

\theoremstyle{definition}

\newcommand{\iid}{\stackrel{\rm iid}{\sim}}
\newcommand{\ind}{\stackrel{\rm ind}{\sim}}

\title{A Bayesian semi-parametric model for longitudinal growth and appetite phenotypes in children}

\author[1]{Andrea Cremaschi}
\author[2]{Beatrice Franzolini}
\author[3,4]{Maria De Iorio}
\author[4,5]{Mary Chong}
\author[4]{Toh Jia Ying}
\author[4]{Navin Michael}
\author[4,6]{Varsha Gupta}
\author[7]{Fabian Yap}
\author[3]{Yung Seng Lee}
\author[3,4]{Johan Erikkson}
\author[4]{Anna Fogel}

\affil[1]{School of Science and Technology, IE University, Spain}
\affil[2]{Institute for Data Science and Analytics, Bocconi University, Italy}
\affil[3]{Department of Paediatrics, Yong Loo Lin School of Medicine, NUS, Singapore}
\affil[4]{Institute for Human Development and Potential (A*STAR), Singapore}
\affil[5]{Saw Swee Hock School of Public Health,  NUS, Singapore}
\affil[6]{Bioinformatics Institute (A*STAR), Singapore}
\affil[7]{Division of Paediatric Medicine, KK Women's and Children's Hospital, Singapore}

\date{}

\begin{document}

\maketitle
	
\section*{Abstract}
	
This study develops a Bayesian semi-parametric model to examine the longitudinal growth and appetite phenotypes in children from the GUSTO cohort, with a focus on understanding the relationship between eating behaviours and growth outcomes over time. While eating behaviours, such as picky eating, have been shown to influence future weight and obesity risk, their developmental patterns and associations with growth trajectories remain under-explored. This work addresses these gaps by modelling longitudinal data, including both growth metrics (e.g., BMI) and eating behaviours (e.g., Child Eating Behaviour Questionnaire, CEBQ), across multiple time points. We extend the Partial Credit Model, commonly used for questionnaire data analysis, to accommodate repeated measurements and incorporate covariates. The growth outcomes are modelled using flexible splines regression. The two components of the model are linked through a shared Bayesian nonparametric prior distribution, specifically a Normalized Generalized Gamma process, allowing to identify clinically relevant subgroups. This joint modelling approach offers a more nuanced understanding of how early eating behaviours relate to growth patterns and developmental outcomes, providing insights into childhood obesity risk.
	
\section{Introduction}

In the study of growth patterns in children, the monitoring of growth biomarkers such as body mass index (BMI) over time is a standard procedure. More recently, attention has been devoted to the study of other early life phenotypic characteristics such as eating behaviours. Eating behaviours acquired in the first years of life are of paramount importance, as they form the foundation of future dietary patterns \citep{da2024longitudinal}. Eating behaviours in early childhood have been previously associated with child weight \citep{mesas2012selected} and  future risk of developing overweight, obesity and related comorbidities \citep{michael2023longitudinal}. 
The majority of evidence on eating behaviours in early childhood comes from cross-sectional studies and single time-point measures in prospective study designs, utilising validated questionnaires \citep{freitas2018appetite}. Little is known about the development of eating behaviour patterns in the first years of life and their concurrent associations with growth trajectories. Moreover, there is mixed evidence on the stability of various eating behaviours in early childhood, which likely varies by the eating behaviour type \citep{derks2019predictors}. For example, picky eating seems to be transient \citep{herle2020eating}, peaking around three years of age and slowly decreasing at about six years of age \citep{taylor2015picky}. 
Most of the studies in this area focus on a small number of data points across time, limiting a comprehensive examination of eating behaviour trajectories over time. Recent data from the ALSPAC cohort examining multiple data points between the ages of 15 months and 10 years identifies differential developmental patterns of over- and under-eating, as well as fussy eating in early childhood. These patterns are further associated with child weight at age 11 years, providing evidence for differential stability of these traits over time, and their importance in predicting future weight status \citep{herle2020eating}. Some shortcomings of the earlier approaches include a limited number of data points, and consequently, focus on the mean scores at distinct time points rather than tracking of patterns over time, which require a larger number of data points and robust sample sizes. 
Modern approaches to questionnaire data analysis facilitate more nuanced and comprehensive analysis to better understand changes in individual traits over time. Child Eating Behaviour Questionnaire \citep[CEBQ,][]{wardle2001development} has been one of the most recognised and commonly used measures to capture eating behaviours from the first months of life. 

The analysis of questionnaire data is usually carried out by first summing the answers to individual questions into summary scores, often by grouping the questions into sub-categories of relevance, and then modelling these scores directly as continuous observations (possibly after square root transformation to improve normality \citep{bartlett1936square}). While this approach facilitates the implementation, it involves a loss of information due to the aggregation of the questionnaire answers, as well as  the assumption made on the support of the data. Alternatively, widely used approaches that do not rely on summary scores are those from Item Response Theory (IRT). IRT is a class of statistical methods belonging to the wider class of latent trait models and employed for modelling questionnaire data, with the aim of capturing a specific \textit{construct} of interest in the survey process. The questions within a questionnaire are commonly referred to as items, and can be answered by the respondents with values on a \textit{likert} scale, i.e. an ordinal scale in which the differences in magnitude between two values are not explicitly known. For instance, the scale $\{bad, good, very\ good\}$ is often associated to the values $\{1, 2, 3\}$, but the differences between consecutive categories do not necessarily represent the same increments in the underlying construct (e.g., depressive symptoms, anxiety traits, appetite phenotypes). In IRT models, the probability of answering within each of the possible categories is modelled via a link function parameterised by question- and subject-specific parameters, thus capturing both the characteristics of the questions (difficulties) as well as the aptitude of the respondent (traits). Two main classes of IRT models are widely used in the literature, based on the way the probabilities of answering within each available category are modelled. A widely used approach models such probabilities via a logistic function, while a second one models the cumulative probabilities first and then defines the likelihood by taking the differences between two adjacent categories. An example of the first one is the Partial Credit model \citep{masters1982rasch, muraki1992generalized}, which is used in this work, while the second class includes the Graded Response model by \cite{samejima1969estimation}. The Partial Credit model offers an advantage in terms of estimation, as in its specification the parameters are algebraically separable, thus allowing the identification of sufficient statistics \citep{andersen1977sufficient, masters1982rasch}. Under the Partial Credit model, the probability of subject $i$ answering question $j$ with category $h$ is given by:
\begin{align}\label{eq:PCMinitial}
\mathbb{P}\left(Y_{ij} = h \mid \theta_i, \alpha_j, \bm \beta_j\right) &= \frac{\exp\left(\alpha_j \left(h\theta_i - \sum_{l = 0}^h \beta_{jl}\right) \right)}{\sum_{h = 0}^{m - 1}\exp\left(\alpha_j \left(h\theta_i -  \sum_{l = 0}^h\beta_{jl}\right) \right)} \\
&h \in \{0, \dots, m-1\}, \quad j = 1, \dots, J, \quad i = 1, \dots, N \nonumber
\end{align}
where $N$ is the number of respondents, $J$ is the number of questions in the questionnaire, and $m > 1$ is the number of possible answering categories, assumed to be the same for all questions. The subject-specific parameter $\theta_i \in \mathbb{R}$ represents the \textit{latent trait}, and captures the answering profile of the $i$-th subject. Within the range of estimated latent trait parameters, higher values of $\theta_i$ indicate a higher propensity of the respondent towards answering with higher values within the answering categories. As such, a higher $\theta_i$ corresponds to a higher intensity of the construct for subject $i$. Eq.~\eqref{eq:PCMinitial} also contains the item-specific parameters $\alpha_j > 0$ and $\bm \beta_j \in \mathbb{R}^m$, for $j = 1, \dots, J$. The first one is a \textit{discrimination parameter}, indicating the ability of the $j$-th question to differentiate between respondent profiles, while $\bm \beta_j$ is a $m$-dimensional vector of \textit{difficulty parameters} for the $j$-th question. In particular, the value $\beta_{jl}$ indicates the increment in construct level required to answer question $j$ with category $l$ rather than $l-1$.

In several clinical settings, as in the one analysed in this work, questionnaires are administered at repeated visits, thus adding a longitudinal dimension to the study. Also in this case, modelling of individual answers is preferred to avoid loss of information due to the use of summary scores \citep{gorter2015item}. This framework is of particular interest for the problem at hand, as it allows to investigate the evolution over time of the appetite phenotypes, via the study of trajectories of latent traits. Extensions of standard IRT models to the longitudinal setting involve the specification of suitable distribution for the latent trait parameters, as well as the estimation of common question-specific parameters throughout the different time points \cite[see, among others,][]{embretson1991multidimensional, andersen1985estimating, von2011measuring, huang2015multilevel, paek2016specifying}. Modelling of time evolution in IRT models can be achieved in different ways. \cite{fischer1995some} and \cite{spiel1998item} propose including time-specific constant effects in the latent trait specification. This modelling strategy is also implemented in the popular R package \texttt{eRm} \citep{mair2016package}. A natural extension is proposed by \cite{verhagen2013longitudinal}, who use a Bayesian latent growth model approach, adequately constrained to avoid identifiability issues. Alternatively, \cite{cremaschi2021bayesian} model the time-varying latent traits via an AR(1) model, also imposing identifiability constraints. Usually the repeated questionnaires contain the same questions across time points, and therefore question-specific parameters are usually shared at different time points. However, this might be generalised in situations where only a subset of the questions is common through the time window of observation \citep{cremaschi2021bayesian}. 

In this work, we develop a joint modelling approach for longitudinal (e.g., BMI scores) and questionnaire (e.g., CEBQ) data, with the aim of investigating their interdependence and the relationship with child development. In particular, we model the longitudinal outcomes via a flexible B-splines regression, while the answers to questionnaires at different time points are modelled using an extension of the Partial Credit model in Eq.~\eqref{eq:PCMinitial} by \citep{masters1982rasch} to allow for answers recorded at different time points and inclusion of covariates. The two components of the model , the BMI trajectories and the CEBQ questionnaires over time, are then connected by the specification of a flexible joint prior distribution over the response-specific parameters. In particular, we use Normalised Generalised Gamma process prior \citep{brix1999generalized, lijoi2007controlling} to induce clustering of the subjects and aid the identification of clinically-relevant subgroups.

The paper is structured as follows. Section 2 introduces the GUSTO data and the model, composed of longitudinal, IRT and Bayesian nonparametric parts; Section 3 presents the results when the proposed method is applied to the GUSTO dataset and Section 4 concludes with a discussion. Additional information regarding the dataset, MCMC algorithm and results is provided in Supplementary Material.
	
\section{Data and Model}
	
The application motivating the proposed approach is the joint study of growth development and the evolution of appetite phenotypes in children. We use data from the Growing Up in Singapore Towards healthy Outcomes (GUSTO) cohort study, a Singaporean prospective birth cohort started in 2009 and still ongoing \citep{soh2014cohort}. GUSTO is a richly phenotyped cohort, containing a large amount of information on mother-child pairs across various clinical domains. Information is available on children's growth trajectories, eating behaviour and mental health; maternal pre- and post-partum characteristics such as physical and mental well-being, socio-demographic variables and medical history. Full details on the GUSTO cohort can be found in \cite{soh2014cohort}.

The main outcomes of interest in this study are the standardised BMI trajectories (Z-BMI) calculated from the child’s height and weight measured during clinical visits at respective time-points described in detail in \cite{soh2014cohort}, and caregiver reported child eating behaviours measured using the CEBQ. Children’s height and weight are recorded at 14 unequally spaced time points from one to nine years of age while eating behaviour is measured via the CEBQ questionnaire at one, three, five and seven years of age. We use WHO guidelines for defining overweight (BMI Z-score $>$ 1 SD) and obesity (BMI Z-score $>$ 2 SD), as well as underweight (BMI Z-score $<$ -2 SD). For ease of explanation, we will refer to children with overweight and/or obesity, as children with overweight. The individual questions of the CEBQ are grouped into eight independent sub-scales representing different eating behaviours, four of which capture what could be described as the children's \textit{Food Approach} (FAp) style behaviours, while the other four capture what could be described as the \textit{Food Avoidance} (FAv) style behaviours.  
Food approach and food avoidance represent two distinct dimensions of children's eating behaviour. Food approach behaviours reflect a heightened interest in food, characterised by an increased appetite and responsiveness to food cues, such as the smell or sight of food. Subscales within this dimension include enjoyment of food, food responsiveness, and emotional overeating, which together may increase the risk of overeating and weight gain. In contrast, food avoidance behaviours involve a reduced interest in eating, selective or picky eating habits, and a tendency to eat slowly or in small amounts. These are captured through subscales like satiety responsiveness, food fussiness, and emotional under-eating, which are often linked to undernutrition or difficulties with maintaining energy intake. 
The questions and sub-scales are reported in Supplementary Table \ref{SMtab:CEBQ_subscales}. We point out that the questionnaires are designed to be answered by the primary caregivers and are aimed at capturing the eating behaviours of the children. Additionally, we include time-homogeneous covariates, listed in Supplementary Table \ref{SMtab:Covariates}.

We now introduce some notation and describe how BMI trajectories and the questionnaire data are modelled. We denote the longitudinal observations by $\bm Z = \left[ Z_{it} \right]$ for $i = 1, \dots, N$ and $t = 1, \dots, T_Z$. The number of mother-child pairs used for the analysis is $N = 282$, while the number of unequally spaced time points of observation for the Z-BMI score is $T_Z = 14$, ranging from one to nine years of age. The time component is modelled via a cubic B-spline $\bm B_Z$ constructed using the R package \texttt{splines} \citep{hastie2017generalized}, where the knots are fixed to the extremes and median values of the observed time points. In this setting, the time evolution of the data is described by a linear combination of polynomial basis terms, with subject-specific coefficients $\bm b_i$. Additionally, fixed covariates $\bm X^Z_i$ are included in the analysis. Thus, for each $i = 1, \dots, N$, the vectors of longitudinal observations $\bm Z_i$ are modelled as multivariate Gaussian with mean vector $\bm b_i B_Z + \bm \gamma^Z \bm X^Z_i \bm 1_{T_Z}$ and diagonal covariance matrix with entries equal to $\sigma^2_Z$, with $\bm \gamma^Z$ a vector of regression coefficients and $\bm 1_{T_Z}$ a vector of ones. For $i = 1, \dots, N$, the longitudinal part of the model is given by: 
\begin{align}\label{eq:Longitudinal_model}
& \bm Z_i \mid \bm b_i, \bm B_Z, \bm \gamma^Z, \bm X^Z_i, \sigma^2_Z \sim \text{N}_{T_Z}\left(\bm b_i \bm B_Z + \bm \gamma^Z \bm X^Z_i \bm 1_{T_Z}, \mathbb{I}_{T_Z} \sigma^2_Z\right) \nonumber \\
& \bm \gamma^Z \sim \text{N}_{q_Z}\left(\bm 0, \mathbb{I}_{q_Z}\right) \\
& \sigma^2_Z \sim \text{IG}\left(3, 2\right) \nonumber
\end{align}
where $\text{N}_d(\bm Y | \bm \mu, \bm \Sigma)$ is the $d$-dimensional Gaussian distribution for the vector $\bm Y$ with mean $\bm \mu$ and covariance matrix $\bm \Sigma$, $\bm 0_{d}$ is a $d$-dimensional vector of zeros and $\mathbb{I}_{d}$ is the $d$-dimensional identity matrix. IG$(a, b)$ indicates the inverse-gamma distribution with mean $b/(a-1)$. 

Let $\bm Y = \left(\bm Y^1, \dots, \bm Y^{T_Y} \right)$ be the array containing the questionnaire data, answered by the participants at $T_Y$ time points. Answers are given by the mothers at one, three, five and seven years of age of the child, so that $T_Y = 4$. The CEBQ questionnaire is composed of $J = 35$ questions with $m = 5$ answer categories ranging from 1 (\textit{Never}) to 5 (\textit{Always}) \citep{wardle2001development}, so that $\bm Y^t = \left[ Y^t_{ij} \right]$, for $i = 1, \dots, N$ and $j = 1, \dots, J$. As reported in Supplementary Table \ref{SMtab:CEBQ_subscales}, the questions are grouped into $n_s = 8$ subscales reflecting the construct to be measured (e.g., EF = \textit{Enjoyment of Food}, FF = \textit{Food Fussiness}). Moreover, the subscales are grouped into $n_p = 2$ main domains (FAp and FAv), introduced earlier, see also Supplementary Table \ref{SMtab:CEBQ_subscales}. We include this information in the model by using suitable indices indicating to which (main) subscale each question belongs. We indicate this by the indices 
$s_j\in\{1,\ldots,n_s\}$ and $p_j\in \{1,\dots,n_p\}$,
 where $s_j$ denotes the subscale and $p_j$ indicates the domain to which each item belongs to, for $j=1,\ldots,J$. We use the subscale information to model a different latent trait for each main domain, thus improving the interpretability of the model. As such, $n_p = 2$ latent traits relative to FAp and FAv are included in the model for each subject. We point out that this is not a multiple-latent trait model, where diverse latent traits are included in the specification of the link function and are estimated using all the questions. Here, each latent trait corresponds to a different set of questions, given by the indices above. Finally, we include covariates in the IRT part of the model directly in the specification of the probability of answering in a certain category. For $i = 1, \dots, N$, the IRT part of the model is specified as follows: 

\begin{small}
\begin{align}\label{eq:IRT_model}
&\mathbb{P}\left(Y^t_{ij} = h \mid \theta^{p_j}_{it}, \alpha_j, \bm \beta_j, \bm \gamma^Y_j, \bm X^Y_i\right) =  \!\begin{aligned}[t]
&\frac{\exp\left(\alpha_{j} \left(h\theta^{p_j}_{it} - \sum_{l = 0}^{h-1} \beta_{jl}\right) + h \bm \gamma^Y_j \bm X^Y_i \right)}{\sum_{h = 0}^{m - 1}\exp\left(\alpha_j \left(h\theta^{p_j}_{it} -  \sum_{l = 0}^{h-1}\beta_{jl}\right) + h \bm \gamma^Y_j \bm X^Y_i \right)}\\
		&h \in \{0, \dots, m-1\}, \quad t = 1, \dots, T_Y, \quad j = 1, \dots, J
\end{aligned} \nonumber \\
& \bm \gamma^Y_j \sim \text{N}_{q_Y}\left(\bm 0_{q_Y}, \mathbb{I}_{q_Y}\right) \\
& \bm \theta^p_i \mid \bm \theta^{p,0}_i \sim \text{N}_{T_Y}\left(\bm \theta^{p,0}_i, \mathbb{I}_{T_Y}\right), \quad t = 1, \dots, T_Y, \quad p = 1, \dots, n_p \nonumber \\	
& \log \left(\alpha_j\right) \mid s_j, \mu_{s_j} \ind \text{N}\left(\mu_{s_j}, 1\right), \quad \mu_{s_j} \iid \text{N}\left(0, 1 \right) \nonumber \\
& \left(\beta_{j2}, \dots, \beta_{jm}\right) \sim \text{N}_{m-1}\left(\bm 0_{m-1}, \mathbb{I}_{m-1}\right), \quad \beta_{j1} = 0, \quad j = 1, \dots, J \nonumber
\end{align}
\end{small}
where $\mu_{s-j}$ is the sub-scale specific prior mean of the $\alpha_j$. To ensure the identifiability of the proposed model, we impose some constraints on the prior specification. Specifically, we fix the variances of regression coefficients and IRT parameters to one. In the case of the parameters $\bm \theta^p_i$, this constraint is necessary to establish the scale of the latent traits in the model \citep{wang2020longitudinal}. We also constrain the difficulty parameters $\beta_{j1} = 0$, for $j = 1, \dots, J$, leaving the remaining $m-1$ terms unconstrained. From a structural equation modelling perspective, this corresponds to fixing one of the loadings to a constant \citep{wang2020longitudinal}. Finally, note that different sets of covariates can be included in the two parts of the model, indicated by $\bm X^Z$ and $\bm X^Y$. Specifically, $\bm X^Y$ is included in the model outside the original contribution to the probability, to avoid being affected by the question-specific discrimination parameters. It is, however, scaled by the factor $h$ for identifiability. Furthermore, note that in the IRT model, the regression coefficients are item-specific, allowing to capture question-specific covariate effects.

A desirable feature of our modelling strategy is the ability to identify risk sub-groups in the population by estimating the clustering structure of the subjects. Specifically, we jointly model the subject-specific parameters in the growth (Z-BMI) and eating behaviours (CEBQ) parts of the model via a nonparametric prior distribution. In the Bayesian framework, this is often achieved by specifying a Dirichlet process \citep[DP,][]{ferguson1973bayesian} prior on the parameters of interest. The DP is a prior distribution defined on the space of probability distributions. An appealing feature of the DP is that it induces a random partition of the subjects. For this reason, and for its analytical tractability, the DP is arguably the most popular prior distribution used in Bayesian nonparametric applications. However, it favours partitions composed of a few large clusters and several small clusters (or singletons), a feature usually referred to as the ``rich-gets-richer'' property. To mitigate such effect, alternative prior distributions can be specified, which allow for more flexible partition estimation. In this work, we employ the Normalised Generalised Gamma (NGG) process \citep{brix1999generalized, lijoi2007controlling}, a generalisation of the DP which favours coarser partitions and fewer singletons, mitigating the``rich-gets-richer'' effect of the DP. In particular, the NGG is characterised by two parameters, denoted here by $\kappa > 0$ and $\sigma \in (0,1)$, and a centring distribution $P_0$. In particular if a random measure $P$ is distributed according to a NGG, then it can be represented as 
\begin{align*}
    P&=\sum_{k\geq 1 } w_k \delta_{\tau_k}
    \\ 
    w_k &= \frac{\mathcal{J}_k}{\sum_{i=1}^\infty \mathcal{J}_i} \\
    \tau_k & \iid P_0 
\end{align*}
where the $\mathcal{J}_k$
are the jump sizes,  $\sum_{i=1}^\infty \mathcal{J}_i< \infty $ almost surely and $P_0$  is a non-atomic probability measure. 
The parameter $\kappa$ plays an analogous role as the mass parameter in the DP, controlling the distribution of the number of clusters, while $\sigma $ controls the sizes of the originated clusters. Specifically, values of $\sigma$ close to 1 increase the number of clusters, while keeping the cluster sizes small. Conversely, when $\sigma$ approaches 0, clusters containing more elements are favoured. These features have been shown to be very useful in mixture modelling and clustering \citep{lijoi2007controlling, argiento2020hierarchical}. When $\sigma \rightarrow 0$, the DP is recovered. We provide below the nonparametric part of the model: 
\begin{align}\label{eq:BNP_model}
& \bm \psi_i = \left(\bm b_i, \bm \theta^{1,0}_i, \dots, \bm \theta^{n_p,0}_i\right) \nonumber \\
& \bm \psi_1, \dots, \bm \psi_N \mid P \iid P, \quad P \sim \text{NGG}\left(\kappa, \sigma, P_0\right) \\
& P_0\left(\bm \psi\right) = \text{N}_d\left(\bm b \mid \bm 0_d, \mathbb{I}_{d}\right) \times
	\text{N}_d\left(\bm \theta^{1,0} \mid \bm 0_{T_Y}, \mathbb{I}_{T_Y}\right) \times \cdots \times 
	\text{N}_d\left(\bm \theta^{n_p,0} \mid \bm 0_{T_Y}, \mathbb{I}_{T_Y}\right)
	 \nonumber
\end{align}

\begin{figure}
\resizebox{\textwidth}{!}{
	\centering
		\tikzset{
			latentnode/.style  ={draw,minimum width=2.5em, shape=circle,thick, black,fill=white},
			visiblenode/.style ={draw, minimum width=2.5em, minimum height=2.5em, shape=rectangle,thick, black,fill=black!20}
		}
			
			\begin{tikzpicture}[auto,thick,node distance=5em]
				\node[visiblenode] (XZ) at (-0.5,0) {$\bm X^{Z}$};
				\node[visiblenode] (XY) at (0.5,0) {$ \bm X^{Y}$};
				\node[visiblenode, left of = XZ] (Z) at (-2,0) {$\bm Z$};
				\node[visiblenode, right of = XY] (Y) at (2,0) {$\bm Y$};
						
				\node[latentnode, left of = Z] (sigmaZ2) {$\sigma^2_Z$};
				\node[latentnode, below of = Z] (bi) {$\bm b$};
				\node[latentnode, below of = Y] (theta) {$\bm \theta$};
				\node[latentnode, right of = Y] (alpha) {$\bm \alpha$};
				\node[latentnode, right of = theta] (beta) {$\bm \beta$};
                
				\node[circle, draw=black, double, fit=(Z) (sigmaZ2) (bi), inner sep=3mm, label={[label distance=0cm, scale = 1.5]90:Z-BMI Trajectories}] (MultNorm) {};
				\node[circle, draw=black, double, fit=(Y) (theta) (alpha) (beta), inner sep=3mm, label={[label distance=0cm, scale = 1.5]90:CEBQ}] (GGM) {};
				\node[ellipse, draw=black, double, inner sep=1mm] (NGG) at (0,-3) {$\mathrm{NGG}\left(\kappa, \sigma, P_0\left(\bm \psi = \left(\bm b, \bm \theta\right) \right) \right)$};

				\draw [->] (XZ) -- node [midway, above] {$\bm \gamma^Z$} (Z);
				\draw [->] (XY) -- node [midway, above] {$\bm \gamma^Y$} (Y);
				\draw [->] (sigmaZ2) -- (Z);
				\draw [->] (bi) -- (Z);
				\draw [->] (theta) -- (Y);
				\draw [->] (alpha) -- (Y);
				\draw [->] (beta) -- (Y);
				\draw [->] (NGG) -- (bi);
				\draw [->] (NGG) -- (theta);
				 						
			\end{tikzpicture}}
		
	\caption{Summary of the proposed model. Circles represent random variables, while squares correspond to observations or fixed hyper-parameters.}
	\label{fig:Model_Figure}
\end{figure}
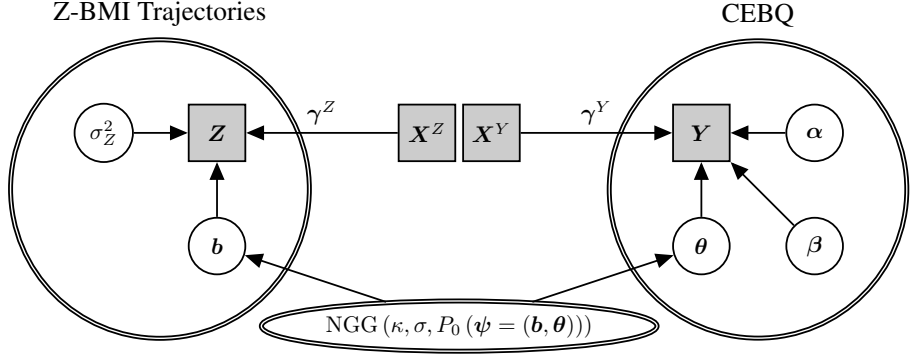

\noindent A schematic representation of the full model is depicted in Figure \ref{fig:Model_Figure}.

\clearpage
\section{Application to the GUSTO data}

We present the results obtained by analysing the GUSTO dataset. We fit the proposed model by fixing the hyperparameters as in Eq.~\eqref{eq:Longitudinal_model} - \eqref{eq:BNP_model}, and the hyperparameters of the NGG process to $\kappa = 1$ and $\sigma = 0.75$. In the application presented in this work, we use different sets of covariates for the longitudinal and IRT parts of the model. In particular, as Z-BMI is computed standardizing for sex, we exclude the covariate \textit{Sex} from $\bm X^Z$. Therefore, after expressing the categorical variables using dummies, the number of covariates is equal to $q_Z = 6$ for the Z-BMI and $q_Y = 7$ for the CEBQ, respectively. The only covariate presenting missing information is the maternal level of education, with 6.38\% missing values, which we impute with the R package \texttt{mice} for multiple imputations \citep{van2011mice}. The missing rate in the Z-BMI observations is below 20\% for all time points and overall equal to 6.51\%, while for the CEBQ questionnaire, it is equal to 0.6\%, 3.51\%, 5.38\% and 0.71\% at each time point, respectively. Imputation of missing responses is carried out naturally in the Bayesian framework by considering them as additional parameters in the model.

We devise an adaptive MCMC Metropolis-within-Gibbs algorithm for the update of the parameters of the proposed model, which is reported in detail in Supplementary Material Section \ref{SMsec:MCMC_Algorithm}. We first run a short burn-in of 100 iterations to initialise the quantities needed in the adaptive MCMC steps. Then, we run 25000 iterations, of which 15000 are discarded as additional burn-in, and the remaining ones are thinned every two yielding a final sample of size 5000.

\clearpage
\paragraph{Clustering}

\begin{figure}[ht]
\centering
\includegraphics[width=1\textwidth]{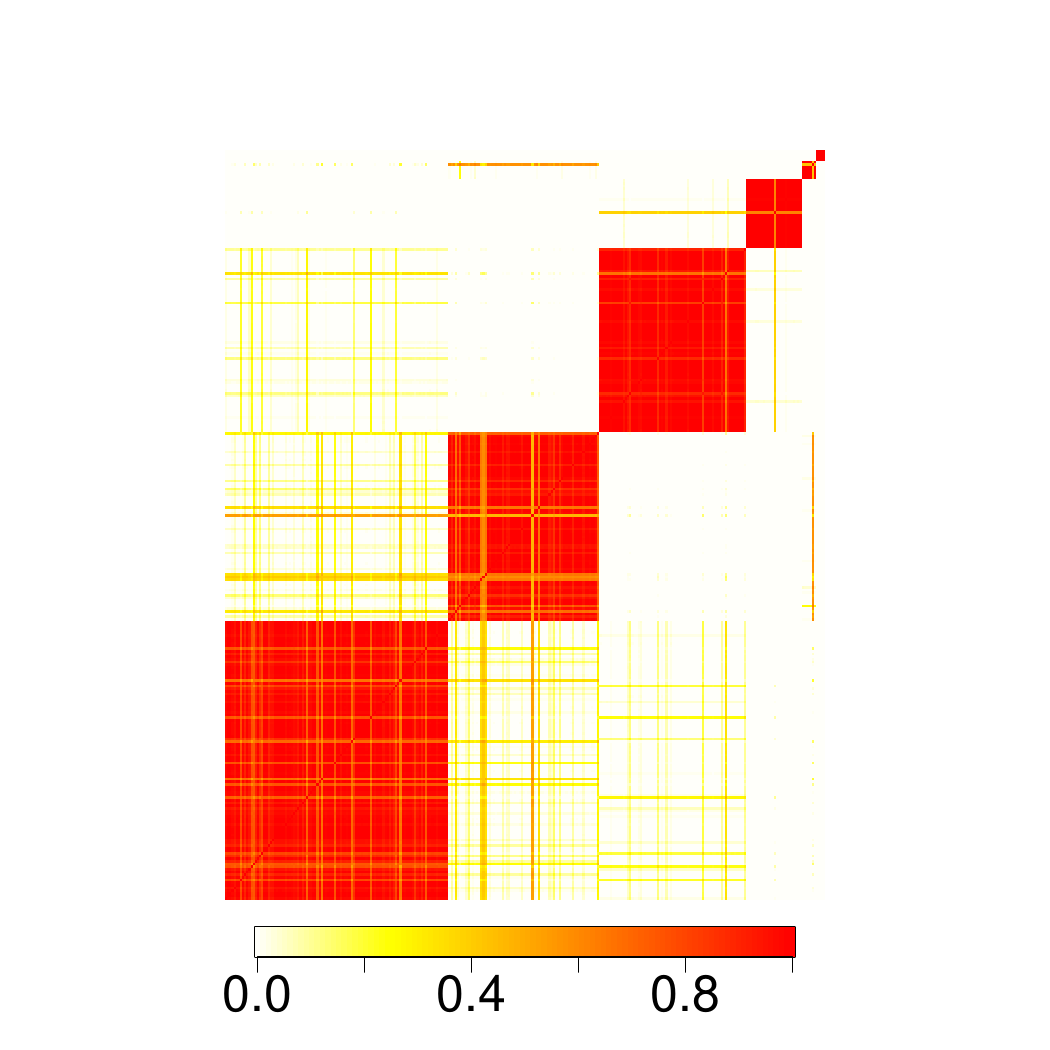}
\caption{GUSTO data. Pairwise posterior co-clustering probabilities, ordered according to the estimated Binder partition.}
\label{fig:CoClustProbs}
\end{figure}

Part~\eqref{eq:BNP_model} of the full model allows the estimate of a partition of the subjects, thanks to the choice of the nonparametric NGG prior used to link subject-specific parameters in the longitudinal and IRT parts. We use the posterior chain of partition samples to estimate the clustering configuration of the subjects. Here, we provide an estimate of the random partition a posteriori by minimising Binder's loss function \citep{binder1978bayesian, lau2007bayesian}, and refer to it as Binder partition. This estimate minimises the expected total misclassification cost due to erroneously clustering two subjects together rather than separately, and vice-versa. The estimated partition also maximises the expected Rand Index \citep{hubert1985comparing, wade2018bayesian}, a measure of similarity between partitions. The estimated Binder partition yields $K_N = 6$ clusters of sizes $\left(105, 71, 69, 26, 7, 4\right)$, where we assign the cluster labels based on their sizes in decreasing order. We show in Figure \ref{fig:CoClustProbs} the posterior co-clustering probabilities ordered according to the estimated Binder partition. We observe a clear pattern in the clustering distribution with little uncertainty.

\begin{figure}[ht]
\centering
\includegraphics[width=1\textwidth]{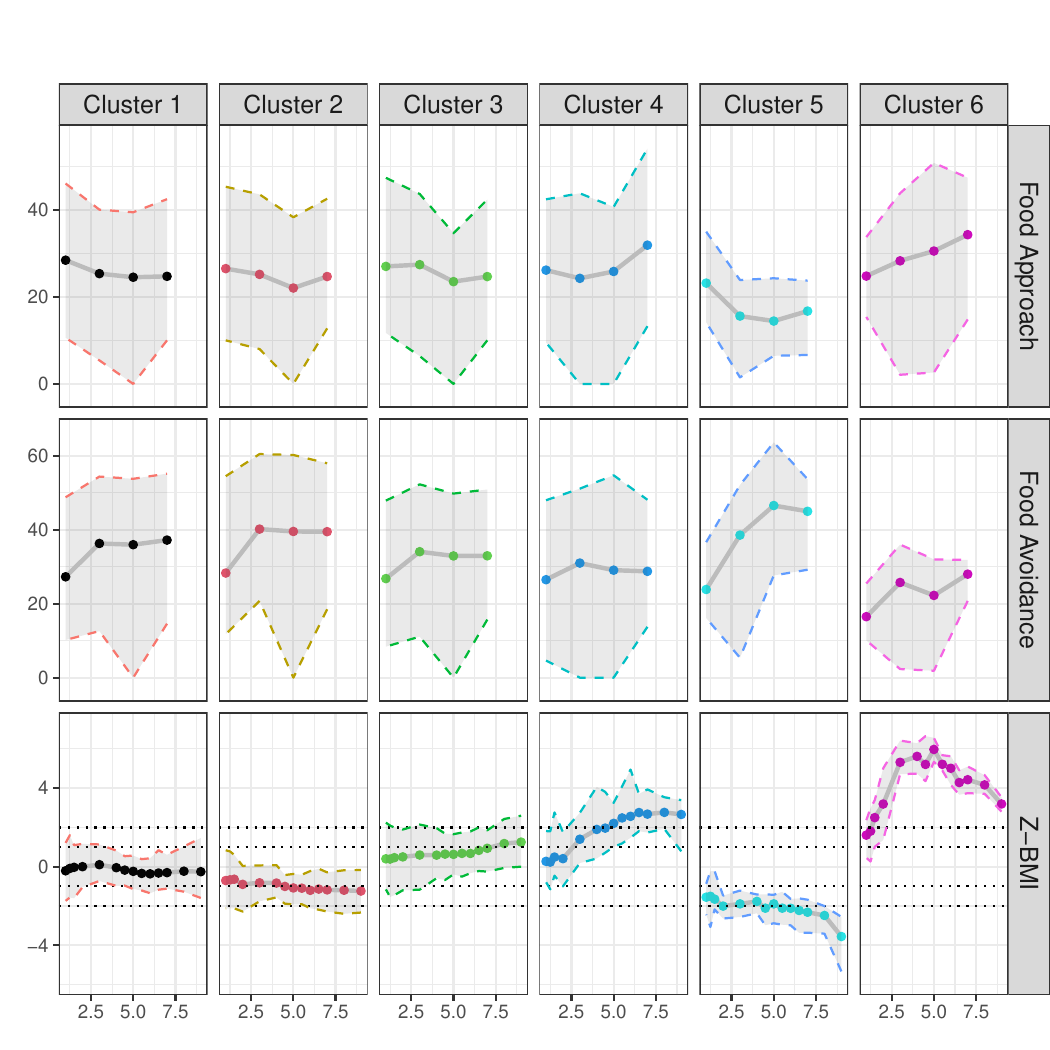}
\caption{GUSTO data. Z-BMI and CEBQ trajectories computed as sample mean and average sum scores, respectively. Shaded regions represent the 95\% empirical quantile range. The first two rows refer to the CEBQ latent traits (Food Approach and Food Avoidance), while the last row reports the average Z-BMI trajectories within each cluster. Dotted lines in the Z-BMI panels indicate the values $\pm$ 1 and $\pm$ 2. Each column refers to a different cluster in the Binder partition. The average trajectories are computed discarding the missing values in the dataset.}
\label{fig:ZY_inClust}
\end{figure}

To characterise the clustering of the subjects, we show in Figure \ref{fig:ZY_inClust} (bottom row) the longitudinal observations (i.e., Z-BMI trajectories) coloured according to the estimated clustering assignment. Note that discontinued lines correspond to missing patterns in the data. The partition of these trajectories suggests a grouping based on different growth patterns, characterised by different magnitudes of Z-BMI trajectories. Additionally, we report in Figure \ref{fig:ZY_inClust} the sample mean trajectories within each cluster and the summary scores (sum of individual answers) obtained within each of the two main subscales of the CEBQ questionnaire, i.e. Food Approach and Food Avoidance. In the IRT part of the model  in Eq.~\eqref{eq:IRT_model}, the sum of the answers to the individual questions is a sufficient statistics for the latent respondent profiles \citep{andersen1977sufficient}. 

Figure~\ref{fig:ZY_inClust} highlights some interesting correspondence between Z-BMI trends and patterns of eating behaviour within the clusters. The three biggest clusters are characterised by stable Z-BMI trajectories, with a slight decrease/increase in later years (approximately after year 3) for Clusters 2 and 3, respectively. These clusters are paired with complementary eating behaviour trajectories, where Cluster 2 has an overall increased Food Approach trajectory compared to Cluster 1, while the opposite can be observed for Cluster 3. Cluster 4 shows a sharp increase in the Food Approach trajectory starting around 2.5 years, accompanied by large values of Z-BMI starting also around the same time point (and plateauing at the end of the time window). This suggests that Cluster 4 groups children who developed overweight/obesity at an early stage, which corresponds to obesogenic eating behaviour patterns at the same time point. Clusters 5 and 6 are the smallest clusters (composed of less than 10 individuals). They present extreme patterns of eating behaviours matched by extremely low/high values of Z-BMI. 




\clearpage
\paragraph{Longitudinal and IRT parameters}

\begin{figure}[ht]
\centering
\includegraphics[width=1\textwidth]{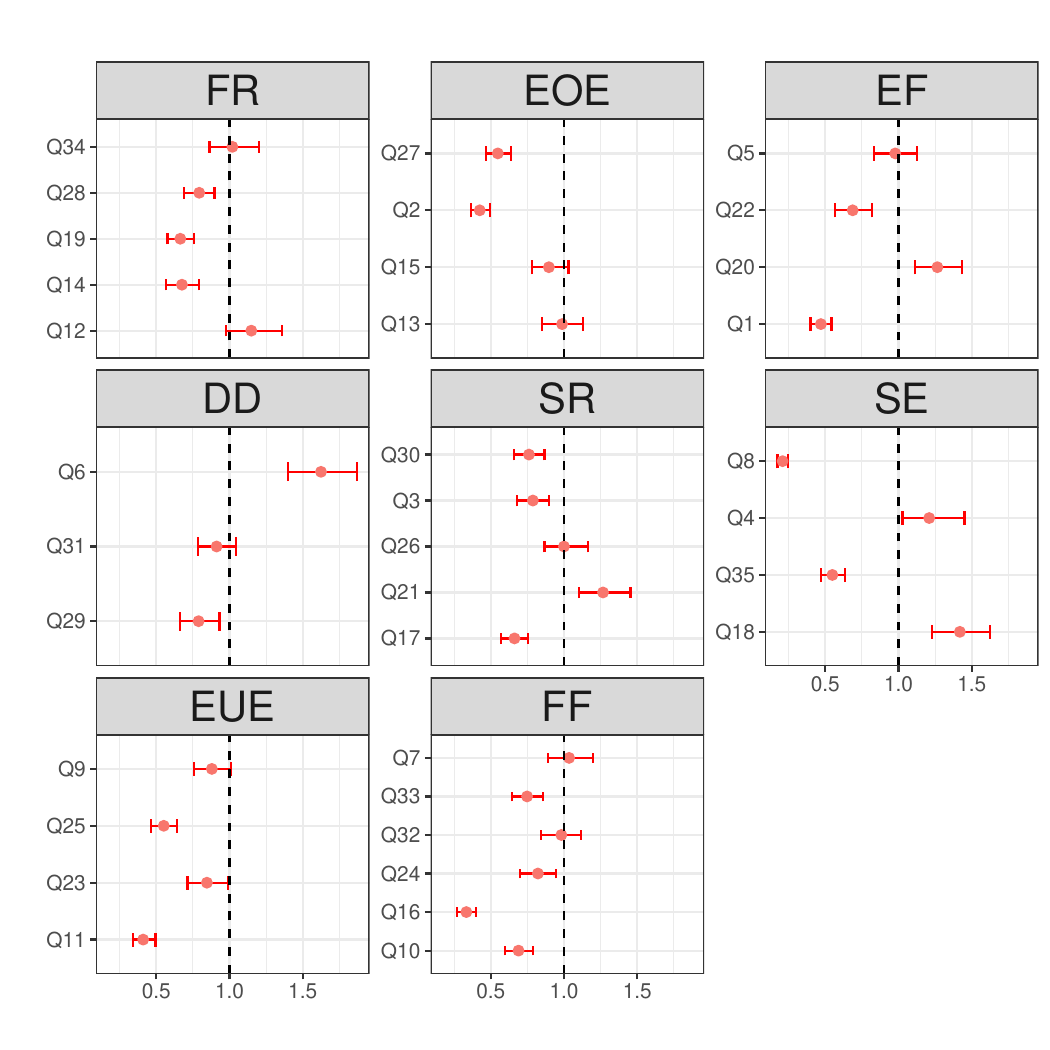}
\caption{GUSTO data. Posterior distribution of the discriminatory parameters $\alpha_j$, for $j = 1, \dots, J$, grouped into subscales.}
\label{fig:ALPHAY_CEBQ}
\end{figure}

Model \eqref{eq:IRT_model} involves some additional parameters which are of interest for this application. In particular, we consider the discriminatory parameters $\alpha_j$ for $j = 1, \dots, J$, characterising the importance of individual questions in differentiating among different response profiles. To highlight this, we show in Figure \ref{fig:ALPHAY_CEBQ} the posterior mean and 95\% credible intervals of these parameters for each question in the CEBQ questionnaire, grouped by subscale. We consider highly discriminatory a question for which the corresponding parameter $\alpha_j$ has a posterior distribution concentrated on values higher than one. From Figure \ref{fig:ALPHAY_CEBQ}, we identify which questions discriminate the most and the least, respectively. We report the five most/least discriminatory questions in Table \ref{tab:Discr_Questions}, obtained by ranking the values of the posterior median of the corresponding discriminatory parameters $\alpha_j$. We observe how the least discriminatory questions have a more extreme construction (``My child always...'') which could encompass a wider variety of situations, or reflect an exaggeration in the maternal response, thus leading to outliers in the type of answers and ultimately to more variability in the estimates. 
Posterior item characteristic curves, i.e. the probability of answering the given question as a function of the latent trait 
$\theta$, relative to these questions are shown in Supplementary Figures \ref{SMfig:ICC_top5} and \ref{SMfig:ICC_btm5}. We can observe how the item characteristic curves relative to more discriminatory questions are more peaked, and the curves relative to answers in the middle (i.e., different from ``Never" and ``Always") place mass on a smaller portion of the support of $\theta$.

\begin{figure}[ht]
\centering
\includegraphics[width=1\textwidth]{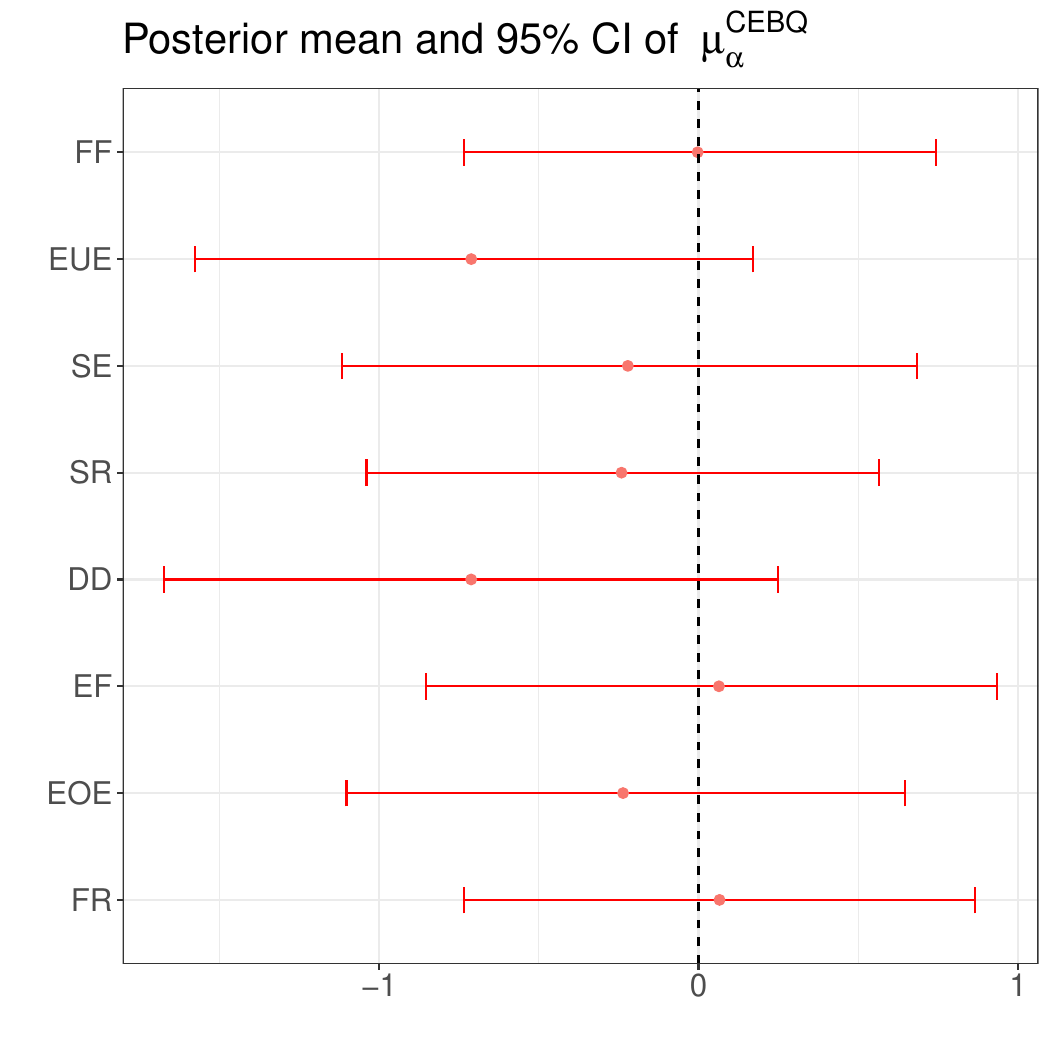}
\caption{GUSTO data. Posterior distribution of the mean parameters $\mu_{s_j}$, for $s_j \in \{1, \dots, n_s\}$.}
\label{fig:MU_ALPHAY_CEBQ}
\end{figure}

The discriminatory parameters in model \eqref{eq:IRT_model} are modelled conditionally to a subscale-specific mean parameters $\mu_{s_j}$, for $s_j \in \{1, \dots, n_s\}$, whose posterior distribution is reported in Figure \ref{fig:MU_ALPHAY_CEBQ}. The posterior mean and 95\% credible interval for the different $\mu_{s_j}$ are concentrated around zero, with lower values obtained for the subscales Emotional Under Eating (EUE) and Desire to Drink (DD), indicating that on average the questions belonging to these subscales are less discriminatory, as it is also observed in Figure \ref{fig:ALPHAY_CEBQ} (with the exception of Question 6).

\begin{table}[h!]
	\caption{GUSTO data. Most and least discriminatory questions,identified by the posterior distribution of the parameters $\alpha_j$, for $j = 1, \dots, J$. The top half of the Table refers to the five most discriminatory questions, while the lowest half to the least discriminatory ones. The corresponding posterior item-characteristic curves are reported in Supplementary Figures \ref{SMfig:ICC_top5} and \ref{SMfig:ICC_btm5}.}
	\label{tab:Discr_Questions}
	\centering
	\begin{tabular}{l|ccl}
 Question \# & main subscale & subscale & Question text\\  \hline
 Q14 & FAp & FR & \textit{If allowed to, my child would eat too much}\\
 Q24 & FAv & FF & \textit{My child is difficult to please with meals}\\
 Q12 & FAp & FR & \textit{My child is always asking for food}\\
 Q19 & FAp & FR & \textit{Given the choice, my child would} \\
  & & & \textit{eat most of the time}\\
 Q22 & FAp & EF & \textit{My child enjoys eating} \\ \hline
 Q11 & FAv & EUE & \textit{My child eats less when s/he is tired}\\
 Q6 & FAp & DD & \textit{My child is always asking for a drink}\\		
 Q29 & FAp & DD & \textit{If given the chance, my child would drink} \\
  & & & \textit{continuously throughout the day}\\ 
 Q31 & FAp & DD & \textit{If given the chance, my child would} \\
  & & & \textit{always be having a drink}\\
 Q23 & FAv & EUE & \textit{My child eats more when she is happy}
 \end{tabular}
\end{table}

\clearpage
\paragraph{Effect of covariates}

\begin{figure}[ht]
\centering
\includegraphics[width=1\textwidth]{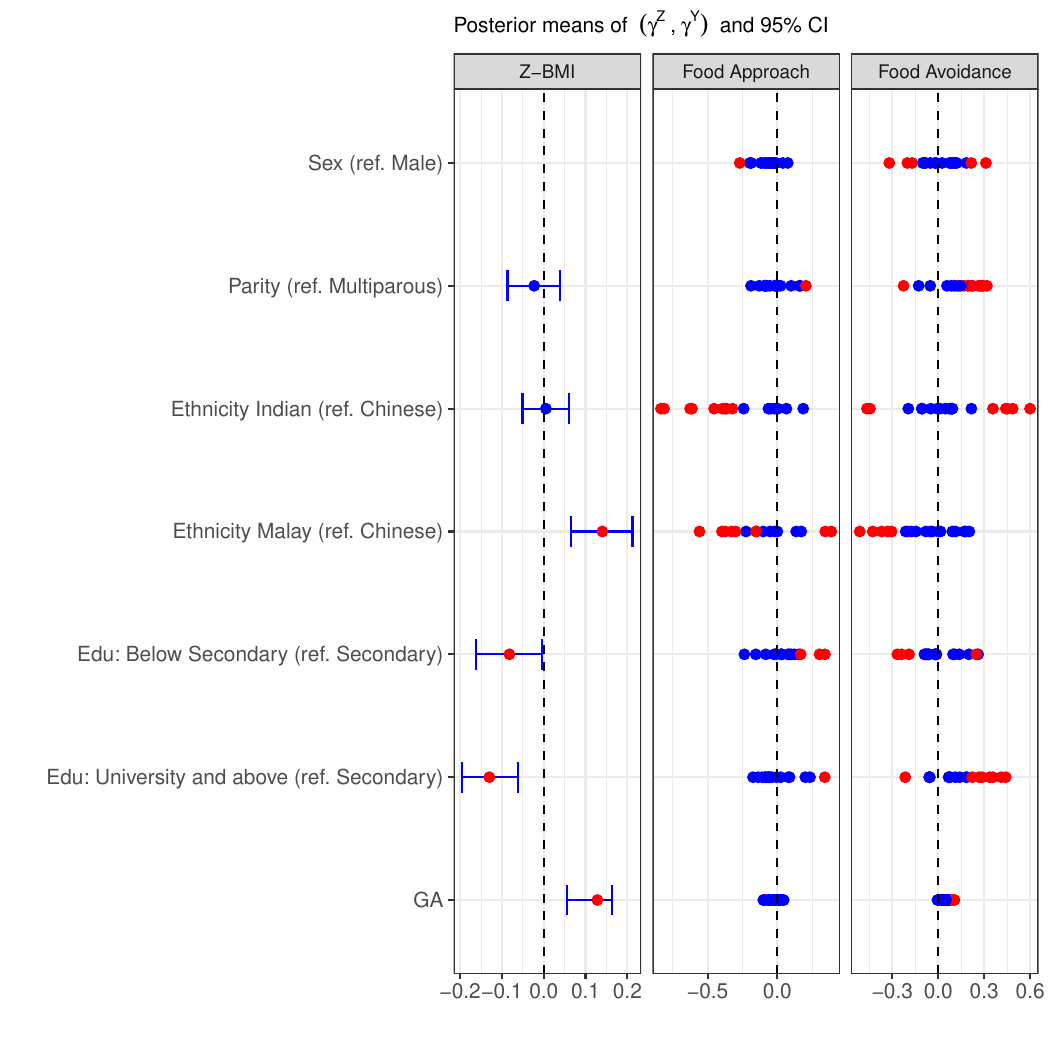}
\caption{GUSTO data. Posterior estimates of the regression coefficients $\bm \gamma^Z$ and $\bm \gamma^Y$ for the fixed covariates. Each panel refer to a different sub-model, i.e. the longitudinal and the questionnaire data. In the latter, the posterior 95\% credibility intervals are omitted for visualisation purposes. Red dots denote regression coefficients whose 95\% credible interval does not contain zero. }
\label{fig:gamma_post}
\end{figure}

We investigate the effect of the fixed-effect covariates on the longitudinal and IRT parts of the model. Recall that the covariate \textit{Sex} is not included in the model for the Z-BMI trajectories, since these are standardised by taking into account sex differences. In Figure \ref{fig:gamma_post}, we show the posterior mean and 95\% credible intervals of the coefficients $\bm \gamma^Z$, while for $\bm \gamma^Y_j$, for $j = 1, \dots, J$, we only show the posterior mean to improve visualisation. Supplementary Figure \ref{SMfig:gamma_post_CEBQ} shows more details. We indicate in red those estimates whose 95\%CI does not contain the value zero, and that are thus considered to have a relevant effect. 

The covariates \textit{Ethnicity}, \textit{Maternal Education} and \textit{Gestational Age} have a stronger effect on the evolution of the Z-BMI trajectories.

The effect of covariates on the respondent profile and appetite behaviour is slightly different when considering the two appetite domains. In particular,  \textit{Sex}, \textit{Parity} and maternal \textit{Education} are characterised by a higher number of significant covariates (positive coefficients) for the questions relative to Food Avoidance. The variable \textit{Ethnicity} has different effects on the two domains corresponding to the different categories: Malay \textit{Ethnicity} has equal negative effects on both domains, while Indian \textit{Ethnicity} has diametrically opposite effects on the two domains. To summarise the results regarding the effects of the covariates on the probability of answering the questions in the CEBQ, we report in Table~\ref{tab:gamma_tab} the number of significant regression effects for each covariate, grouped according to the CEBQ subscales.

\begin{table}[h!]
	\caption{GUSTO data. Number of regression coefficients whose posterior 95\% credible interval does not contain the value zero, and are therefore considered as significant effects.}
	\label{tab:gamma_tab}
	\centering
	\begin{tabular}{l|cccccccc}
 \diagbox{Covariate}{Subscale} & FR & EOE & EF & DD & SR & SE & EUE & FF \\  \hline
 GA & 0 & 0 & 0 & 1 & 0 & 0 & 1 & 1 \\
Edu: Uni and above & 1 & 0 & 0 & 0 & 0 & 1 & 4 & 3 \\
Edu: below Secondary & 1 & 0 & 0 & 2 & 2 & 0 & 1 & 1 \\
Ethnicity Malay & 4 & 1 & 1 & 2 & 1 & 2 & 3 & 0 \\
Ethnicity Indian & 4 & 1 & 4 & 0 & 5 & 1 & 1 & 0 \\
Parity & 1 & 0 & 0 & 0 & 4 & 4 & 2 & 1 \\
Sex (ref. Male) & 0 & 0 & 1 & 0 & 1 & 2 & 2 & 0
 \end{tabular}
\end{table}

\clearpage
\section{Discussion}
	
We identify six distinct trajectories of Z-BMI in early childhood, which are further linked to appetite phenotypes describing food approach and food avoidance style behaviours. Three of the trajectories describe children with healthy BMI. The remaining trajectories describe children who have healthy weight in the first years of life and develop overweight/obesity when they reach school age (Cluster 4), children with low body weight who by school age are classified as underweight (Cluster 5) and one trajectory (Cluster 6) of children who are overweight in the first years of life and develop obesity by school age. 

These six trajectories map on surprisingly well to eating behaviours in the expected direction. Children with healthy weight trajectories (Cluster 1-3) show increased food avoidance behaviours over time, and plateaued-to-decreased food approach behaviours over time. Interestingly, our data suggest that food avoidance behaviours have a stronger influence on a child’s weight compared to food approach behaviours. Among the three clusters of healthy weight, children from Cluster 3 show an increase in BMI over time, albeit still within the healthy ranges, corresponding to a lower increase in food avoidance behaviours as compared to Clusters 1 and 2. No visual differences are detected in food approach behaviours between these three clusters, suggesting that food avoidance may be a stronger determinant of weight. 

Children from Cluster 4 who become overweight/obese show an increase in food approach behaviours over time, which may be driving increased food consumption, supposedly exceeding nutritional requirements and consequently rapid growth in weight. Food approach behaviours such as high emotional overeating, food responsiveness or enjoyment of food have been previously linked to higher BMI in other samples (Wardle et al. 2001; Jaarsveld et al. 2009). Children in Cluster 4 also have higher mean food approach behaviours by school age, compared to children from Clusters 1-3 or Cluster 5 (underweight). At the same time, compared to children with healthy weight trajectories (Clusters 1-3), this group of children shows a smaller concurrent increase in food avoidance behaviours, which remained largely unchanged throughout childhood. Conversely, children with healthy weight trajectories show an increase in food avoidance, and a minor decrease in food approach behaviours,  throughout early childhood. Food avoidance behaviours such as fussy eating, slowness in eating or satiety responsiveness are typically linked with picky eating, food disgust, and low energy intakes and tend to be lower among overweight children compared to those with healthy weight (Galloway et al. 2003; Carnell and Wardle, 2008). This is further seen in Cluster 6 representing children with obesity, whose food avoidance scores are approximately 50\% lower at age 12 months compared to children from Clusters 1-3, and only reach the baseline of children with healthy weight by school age. At the same time, children with obesity show a rapid and dramatic increase in food approach behaviours not observed in any other Cluster. Interestingly, food approach behaviours in children from Cluster 6 are the lowest at age 12 months, constituting the highest recorded scores for underweight children (Cluster 5). Children in Cluster 5 (underweight trajectory) show a rapid and dramatic decrease in food approach behaviours between the ages 1 year to 3 years and a rapid and dramatic increase in food avoidance behaviours between ages 1 year and  5 years, demonstrating a very unique pattern of eating behaviours compared to the other Clusters.
The results of this study suggest that eating behaviours map very well on child growth patterns and that food avoidance behaviours are a stronger differentiating factor compared to food approach behaviours. This study provides a unique set of analyses, where eating behaviours subscales are grouped together into latent variables rather than analysed individually and independently. Reducing the subscales to two latent variables makes the model more parsimonious and easier to interpret, particularly in light of the good correspondence of these variables with child BMI over time, in the expected direction. In addition, a unique analysis of the discriminatory power of individual items highlighted 5 items which are poor differentiators, which belong to two subscales. Interestingly, all three items that belong to Desire to Drink subscale were poor differentiators and two out of the total of 4 items from the Emotional Undereating subscale (marking 50\% of total items). Desire to Drink and Emotional Undereating may not be good differentiators between the different eating behaviour patterns, suggesting potential redundancy of these specific items, or perhaps even the entire subscales, in this specific sample. Nevertheless, the analysis of the latent characteristics reduced the granularity of the importance of these items for the overall analysis.

\clearpage
\bibliographystyle{plainnat}
\bibliography{Biblio}

\end{document}


\title{A Bayesian semi-parametric model for longitudinal growth and appetite phenotypes in children \\ Supplementary Material}
    \date{}
	
	\maketitle
	
	\section*{Abstract}
	Supplementary Material file for the manuscript ``A Bayesian semi-parametric model for longitudinal growth and appetite phenotypes in children''. This file contains additional Figures and Tables referenced in the main manuscript in Section \ref{SMsec:additional-figures-and-tables}, and details on the MCMC algorithm in Section \ref{SMsec:MCMC_Algorithm}.

	\section{Additional Figures and Tables}\label{SMsec:additional-figures-and-tables}
	
	We report in this Section additional Figures and Tables referenced in the main text.

	\begin{table}[!ht]
		\centering
		\caption[]{GUSTO cohort study: Fixed covariates used in the analysis. For categorical covariates, the reference categories are indicated by (ref).}
		\label{SMtab:Covariates}
		\begin{tabular}{c|cc}
			\textbf{Variable name} & \textbf{Levels/Range} & \textbf{Missing $\%$} \\\hline
			
			mother highest education & \makecell{Secondary (ref), \\ Below Secondary, \\ University and above} & 6.38 \\\cline{2-2}
			
			mother ethnicity & \makecell{Chinese (ref), Malay, Indian} & 0.00 \\\cline{2-2} 

			Gestational age & $\mathbb{R}^+$ & 0.00 \\\cline{2-2}
			
			Sex of the child & {Male (ref), Female} & 0.00 \\\cline{2-2}
			
			parity & \makecell{Multiparous (ref), Nulliparous} & 0.00
			
		\end{tabular}
	\end{table}

	\begin{landscape}
	\begin{table}[!ht]
		\centering
		\caption[]{GUSTO cohort study: CEBQ subscales structure \citep{wardle2001development}.}
		\label{SMtab:CEBQ_subscales}
		\begin{tabular}{c|c|l}
			\textbf{Main subscale} & \textbf{Subscale} & \textbf{Question} \\\hline
			
			\multirow{16}{*}{Food Approach (FAp)} 
			& \multirow{5}{*}{Food Responsiveness (FR)}
			& My child is always asking for food \\
			& & If allowed to, my child would eat too much \\	
			& & Even if my child is full up s/he finds room to eat his/her favourite food \\
			& & If given the chance, my child would always have food in his/her mouth \\
			& & Given the choice, my child would eat most of the time \\		

 			& \multirow{4}{*}{Emotional Over-Eating (EOE)}
			& My child eats more when annoyed \\
			& & My child eats more when worried \\
			& & My child eats more when anxious \\
			& & My child eats more when s/he has nothing else to do \\
									
 			& \multirow{4}{*}{Enjoyment of Food (EF)}
 			& My child loves food \\
			& & My child is interested in food \\
			& & My child looks forward to mealtimes \\
			& & My child enjoys eating \\

 			& \multirow{3}{*}{Desire to Drink (DD)}
			& If given the chance, my child would drink continuously throughout the day \\		
			& & If given the chance, my child would always be having a drink \\			
			& & My child is always asking for a drink \\\hline
			
			\multirow{19}{*}{Food Avoidance (FAv)}
			& \multirow{5}{*}{Satiety Responsiveness (SR)}				
			& My child has a big appetite \\
			& & My child leaves food on his/her plate at the end of a meal \\
			& & My child gets full before his/her meal is finished \\
			& & My child gets full up easily \\
			& & My child cannot eat a meal if s/he has had a snack just before \\
			
			& \multirow{4}{*}{Slowness in Eating (SE)}								
			& My child finishes his/her meal quickly \\
			& & My child eats slowly \\
			& & My child takes more than 30 minutes to finish a meal \\
			& & My child eats more and more slowly during the course of a meal \\
			
			& \multirow{4}{*}{Emotional Under-Eating (EUE)}			
			& My child eats less when angry \\
			& & My child eats less when s/he is tired \\
			& & My child eats more when she is happy \\
			& & My child eats less when upset \\
					
			& \multirow{6}{*}{Food Fussiness (FF)}							
			& My child refuses new foods at first \\
			& & My child enjoys tasting new foods \\
 			& & My child enjoys a wide variety of foods \\
			& & My child is difficult to please with meals \\
			& & My child is interested in tasting food s/he hasn't tasted before \\
			& & My child decides that s/he doesn't like a food, even without tasting it

		\end{tabular}
	\end{table}
	\end{landscape}

\begin{figure}[ht]
\centering
\includegraphics[width=1\textwidth]{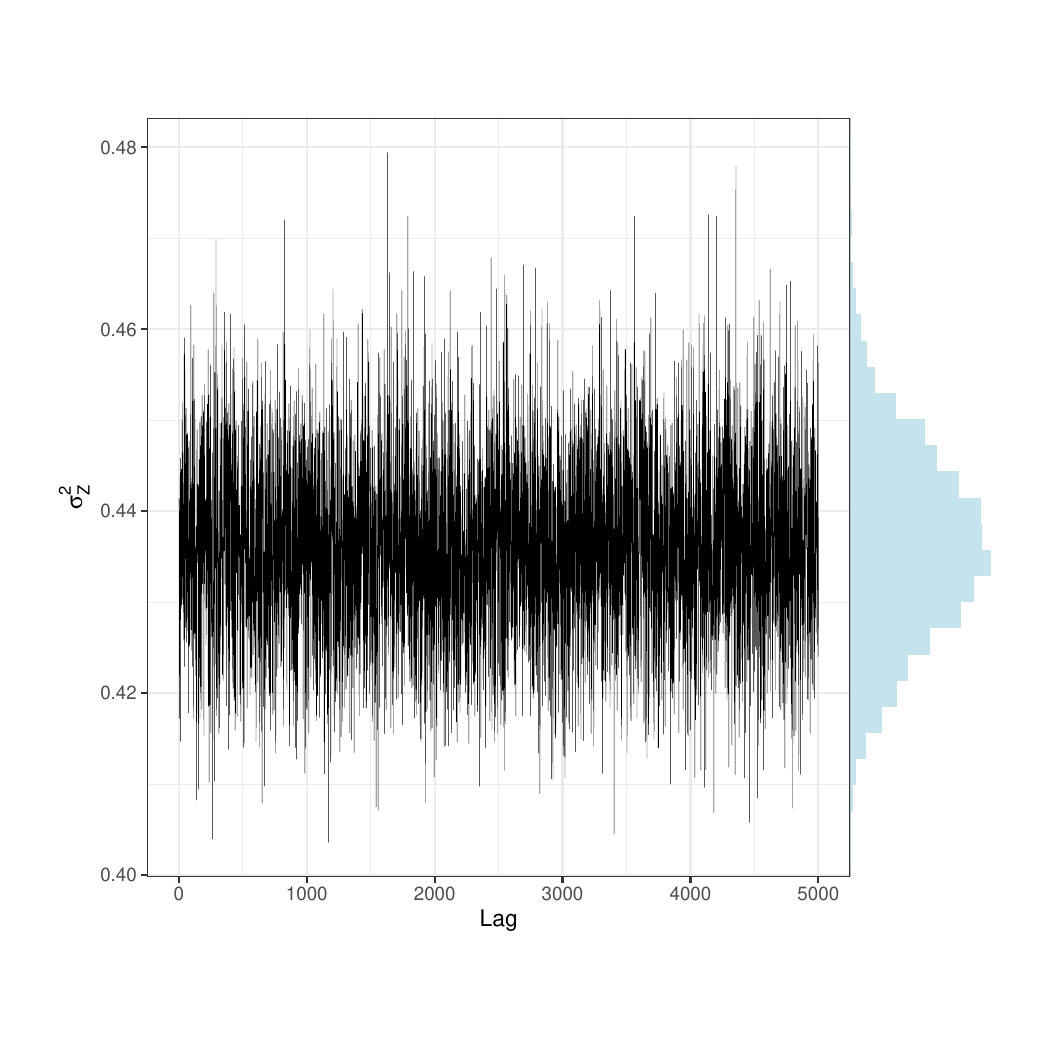}
\caption{GUSTO data. Posterior distribution and MCMC sample of the variance parameter $\sigma^2_Z$.}
\label{SMfig:sig2_Z_post}
\end{figure}

We show in Figure \ref{SMfig:sig2_Z_post} the posterior distribution of the variance parameter $\sigma^2_Z$, capturing the variability around the longitudinal Z-BMI trajectories. This parameter is assumed fixed as it controls the overall variability in the sample. The traceplot of the posterior MCMC chain, also shown in Figure \ref{SMfig:sig2_Z_post}, indicates good convergence.

\begin{figure}[ht]
\centering
\includegraphics[width=1\textwidth]{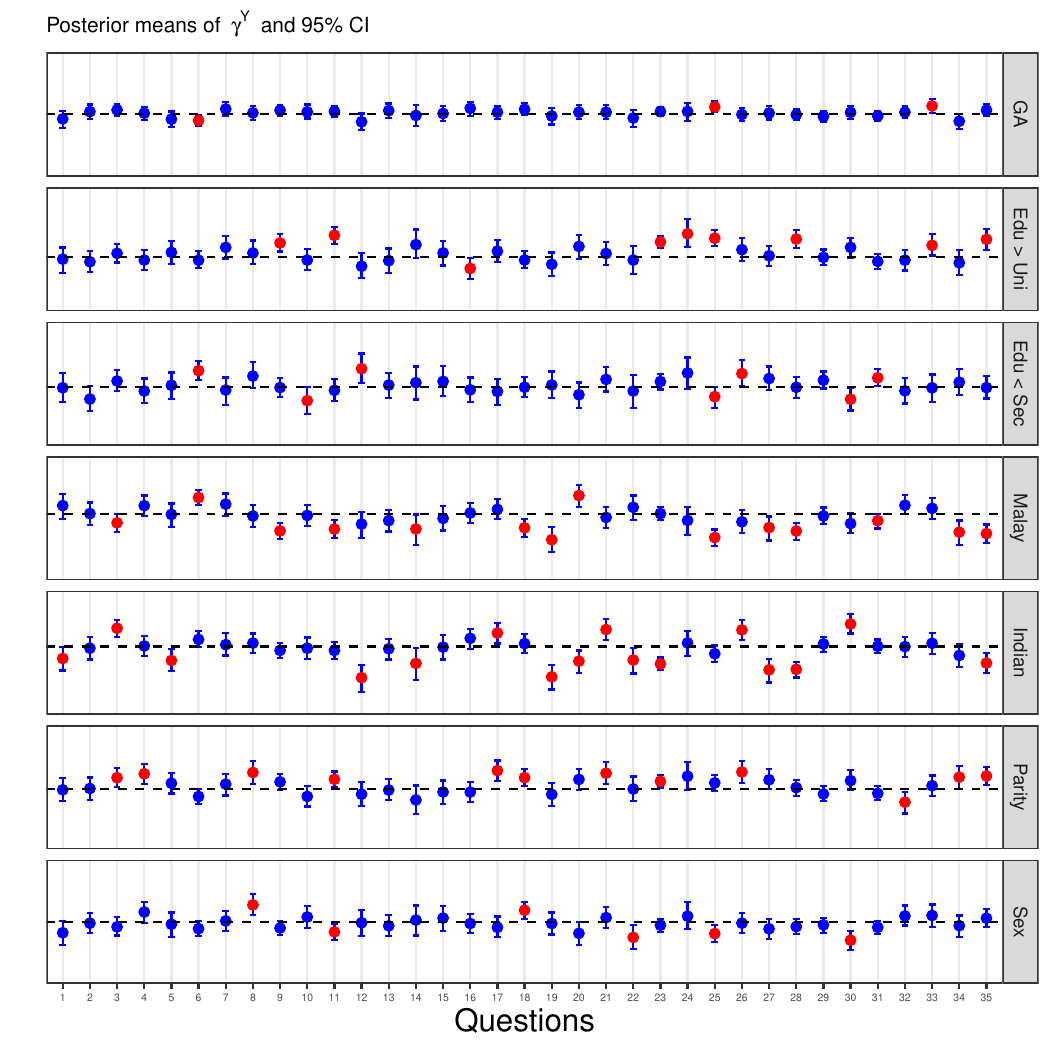}
\caption{GUSTO data. Posterior estimates of the question-specific regression coefficients $\bm \gamma^Y_j$ for $j = 1, \dots, J$. Each row refers to a different covariate. The posterior means of the question-specific regression coefficients, indicated as dots, are coloured in red if the corresponding 95\% credible interval does not contain the value zero.}
\label{SMfig:gamma_post_CEBQ}
\end{figure}

Figure \ref{SMfig:gamma_post_CEBQ} shows the posterior estimates of the regression coefficients $\bm \gamma^Y$, which are question-specific, and their 95\% credible intervals.

\begin{figure}[ht]
\centering
\subfloat[Q4]{\includegraphics[width=0.5\textwidth]{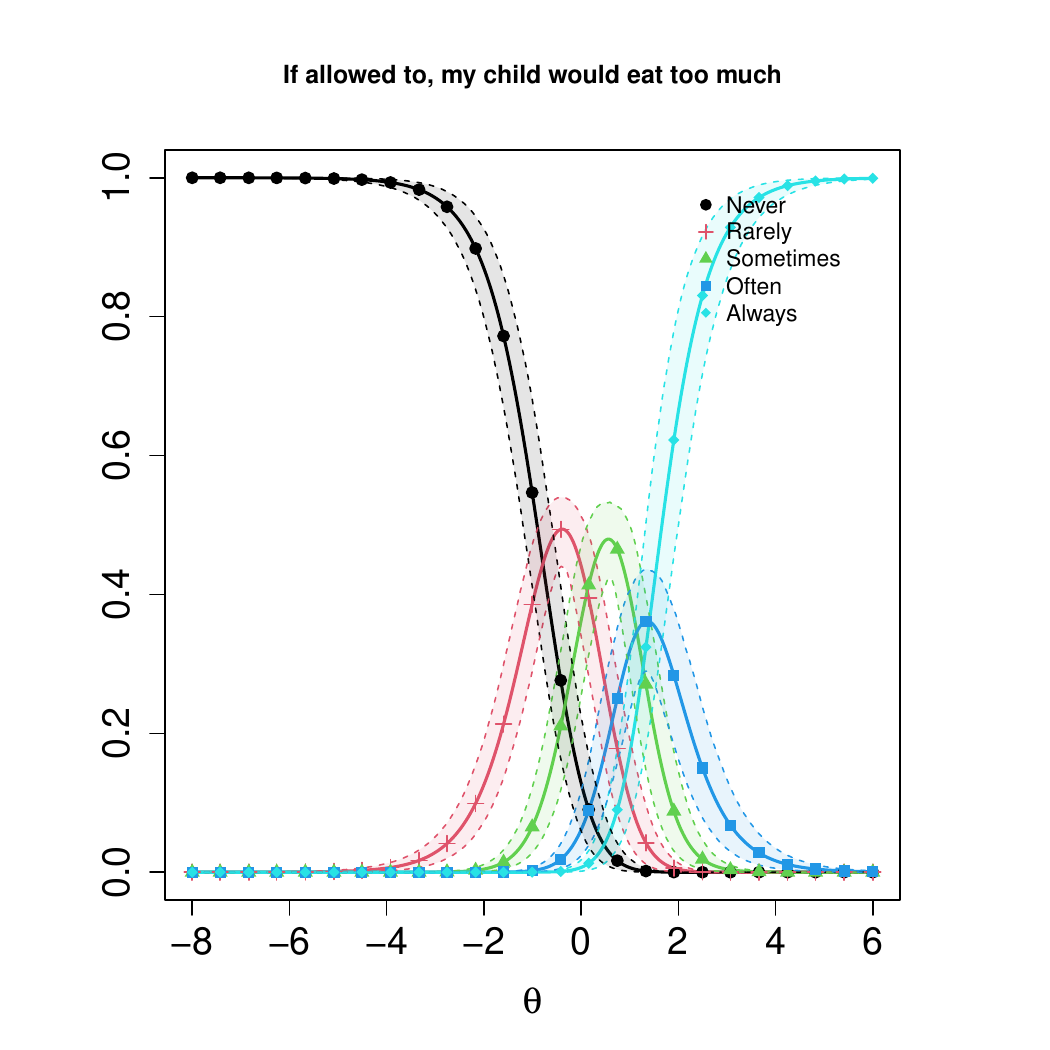}}
\subfloat[Q18]{\includegraphics[width=0.5\textwidth]{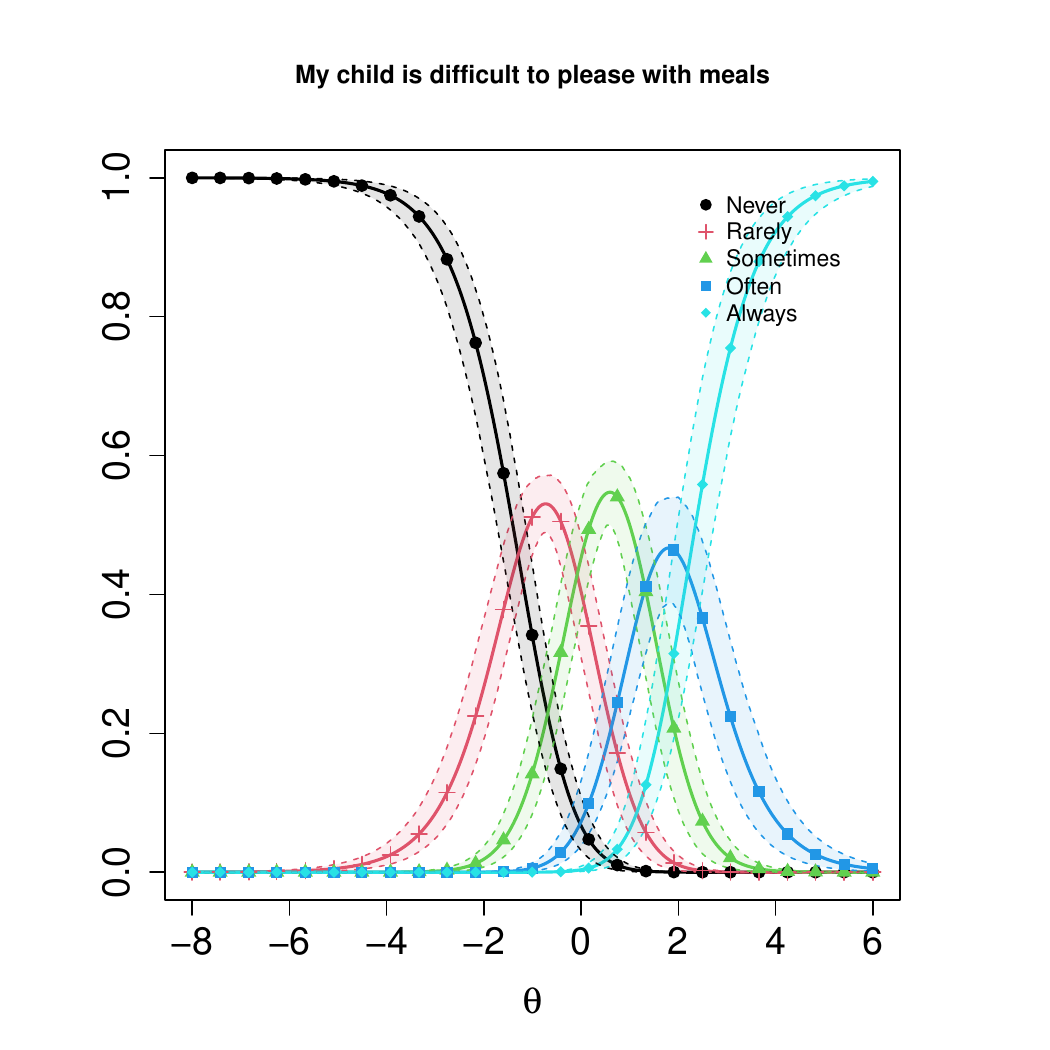}} \\
\subfloat[Q6]{\includegraphics[width=0.5\textwidth]{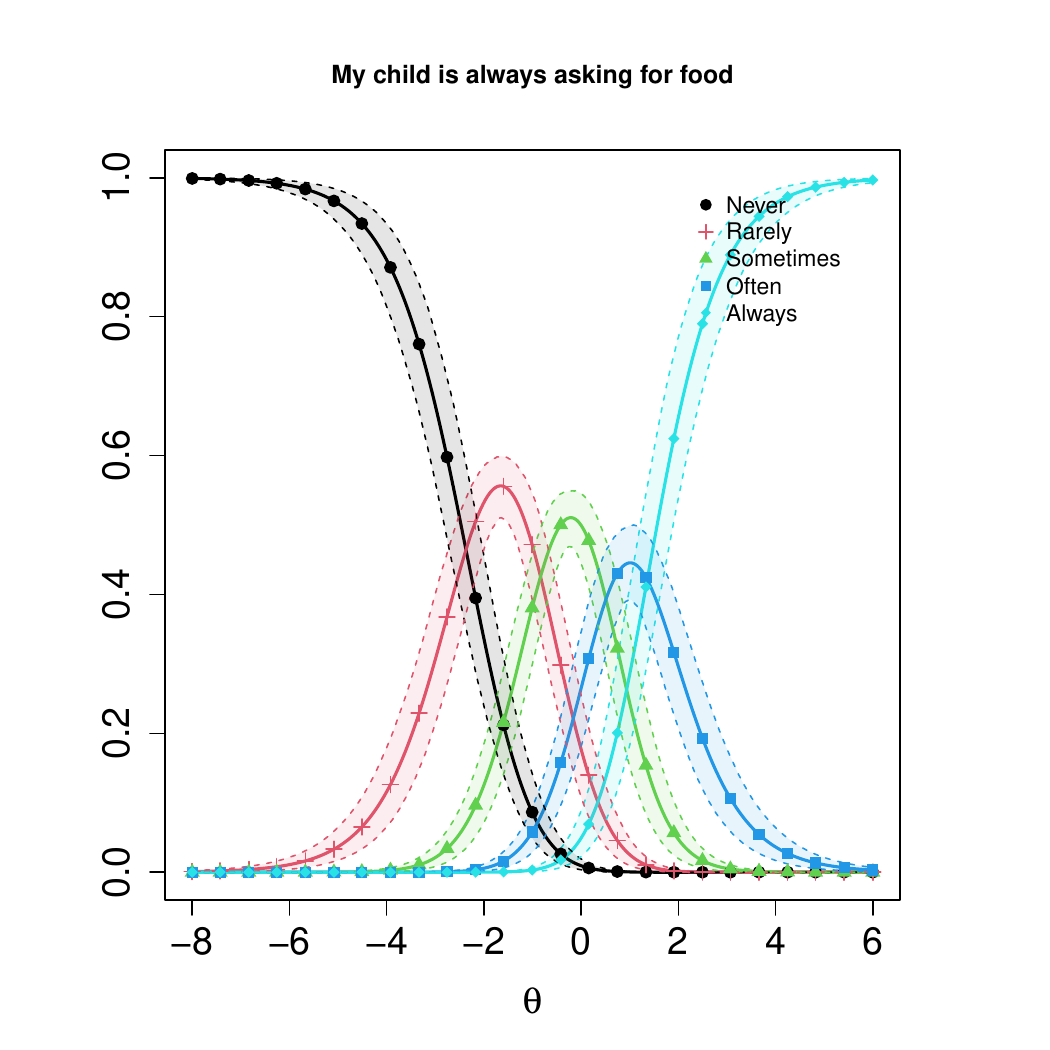}}
\subfloat[Q20]{\includegraphics[width=0.5\textwidth]{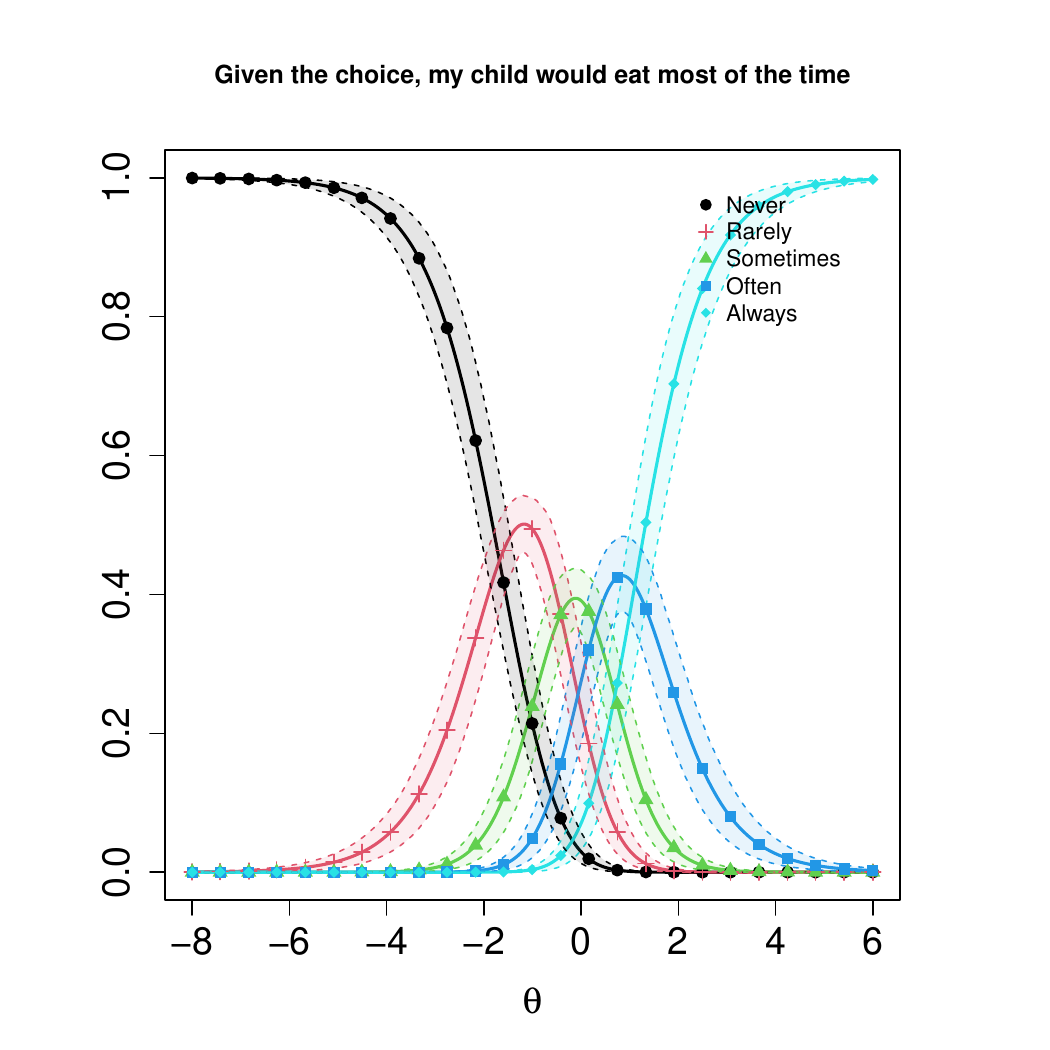}}\\
\subfloat[Q21]{\includegraphics[width=0.5\textwidth]{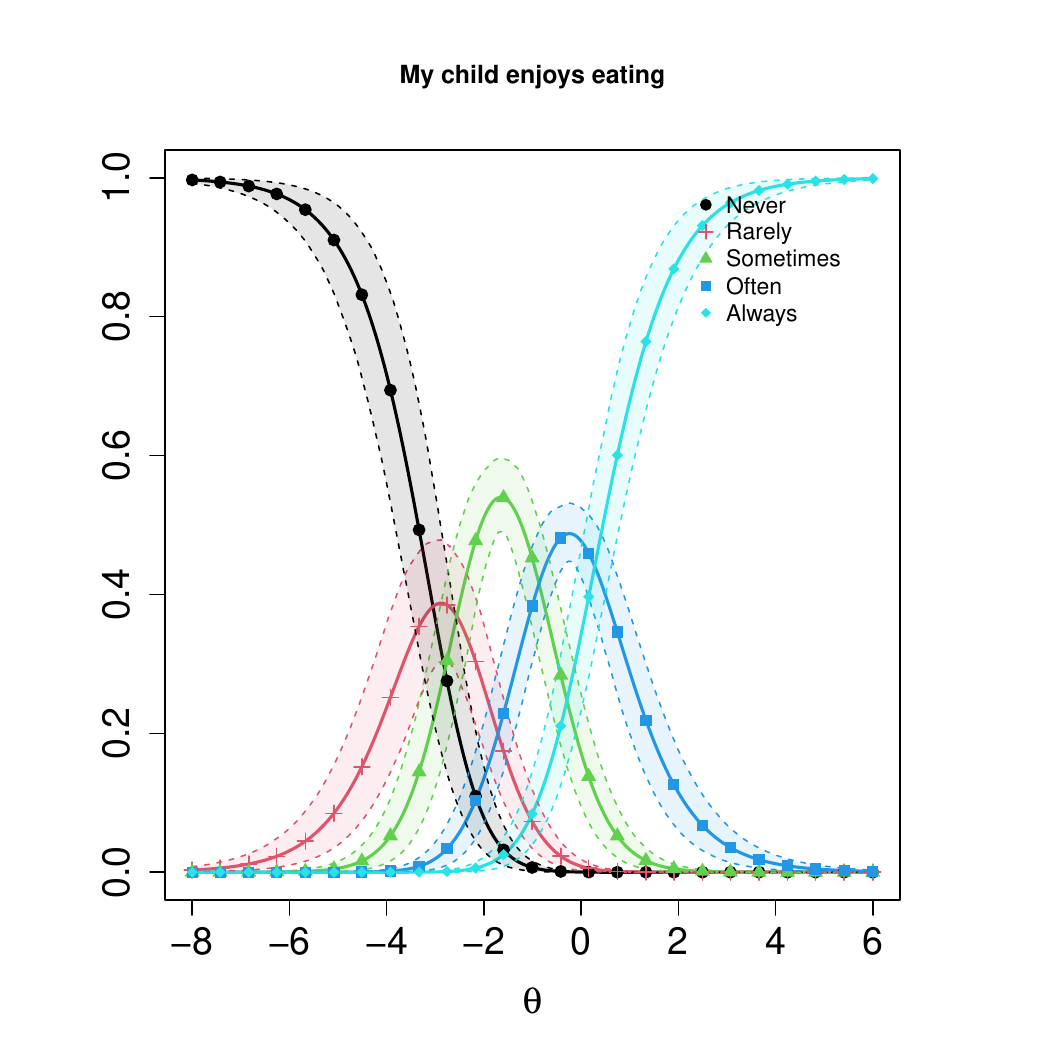}}
\caption{GUSTO data. Posterior item characteristic curves of the five most discriminatory questions, based on the posterior distribution of the corresponding discriminatory parameters $\alpha_j$.}
\label{SMfig:ICC_top5}
\end{figure}	

\begin{figure}[ht]
\centering
\subfloat[Q1]{\includegraphics[width=0.5\textwidth]{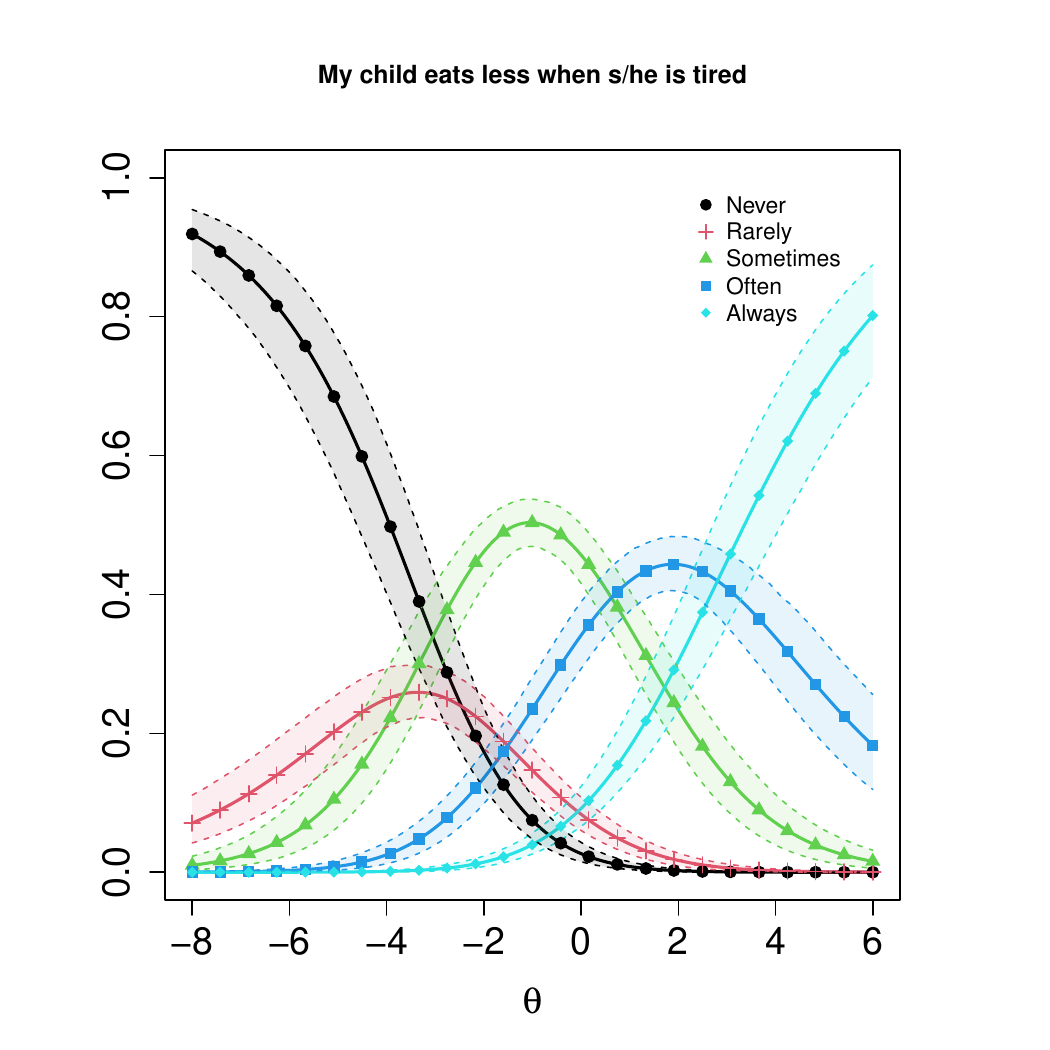}}
\subfloat[Q8]{\includegraphics[width=0.5\textwidth]{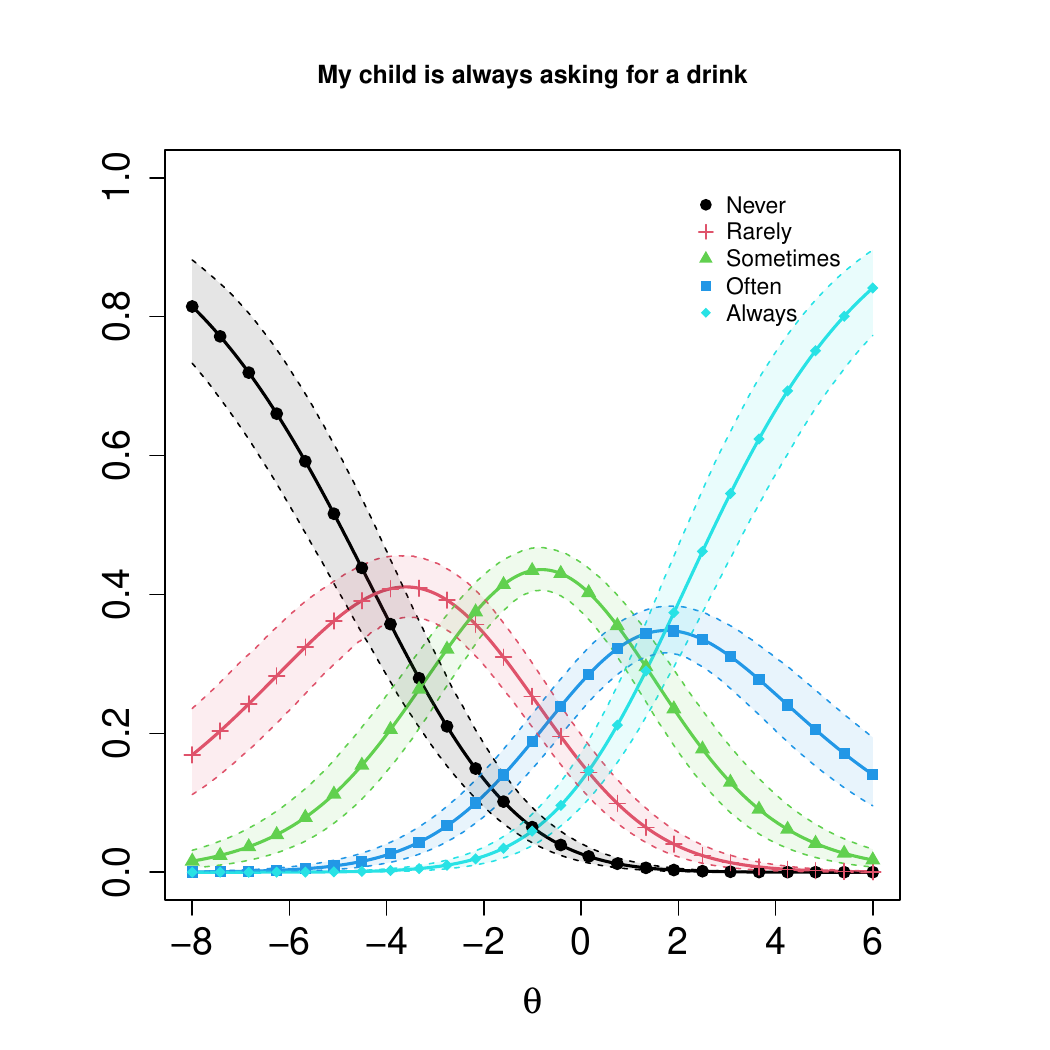}} \\
\subfloat[Q11]{\includegraphics[width=0.5\textwidth]{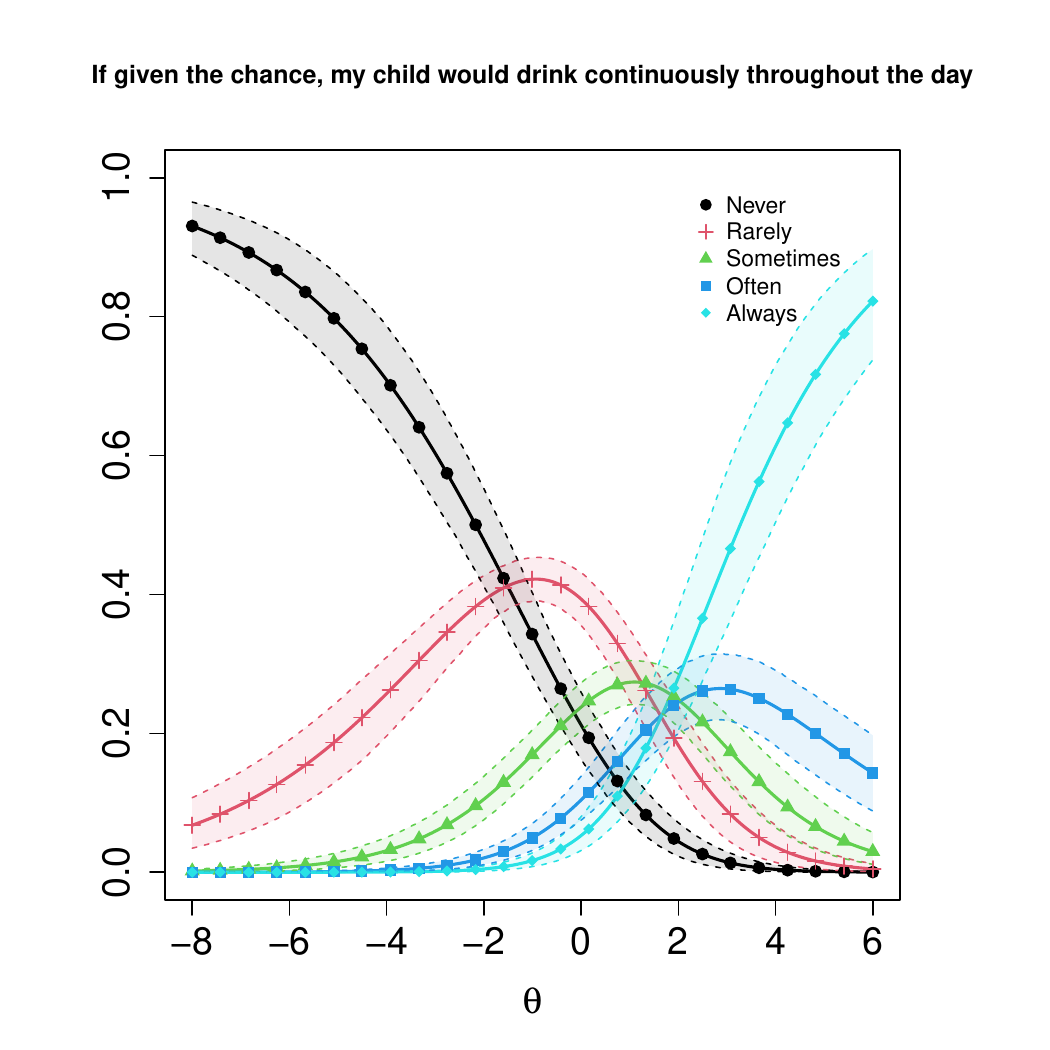}}
\subfloat[Q16]{\includegraphics[width=0.5\textwidth]{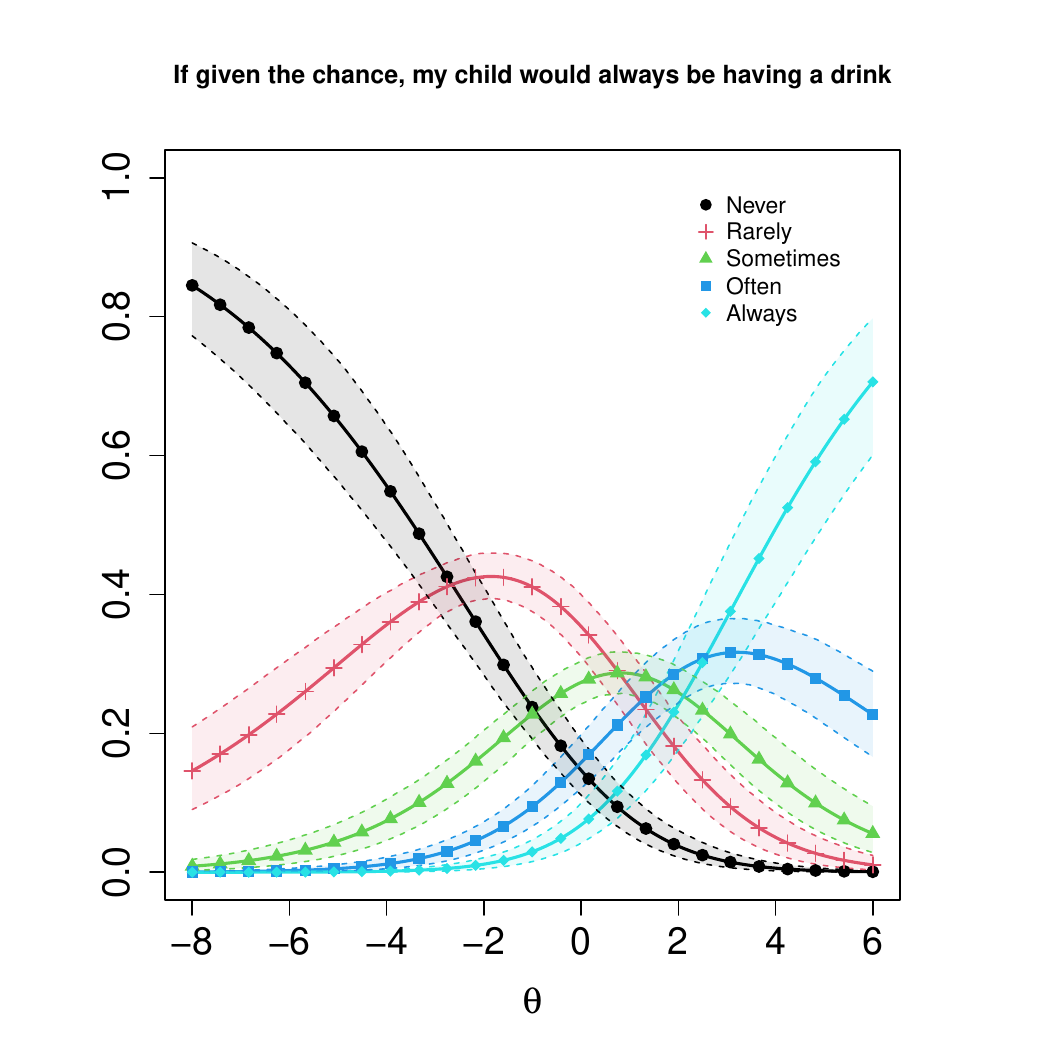}}\\
\subfloat[Q25]{\includegraphics[width=0.5\textwidth]{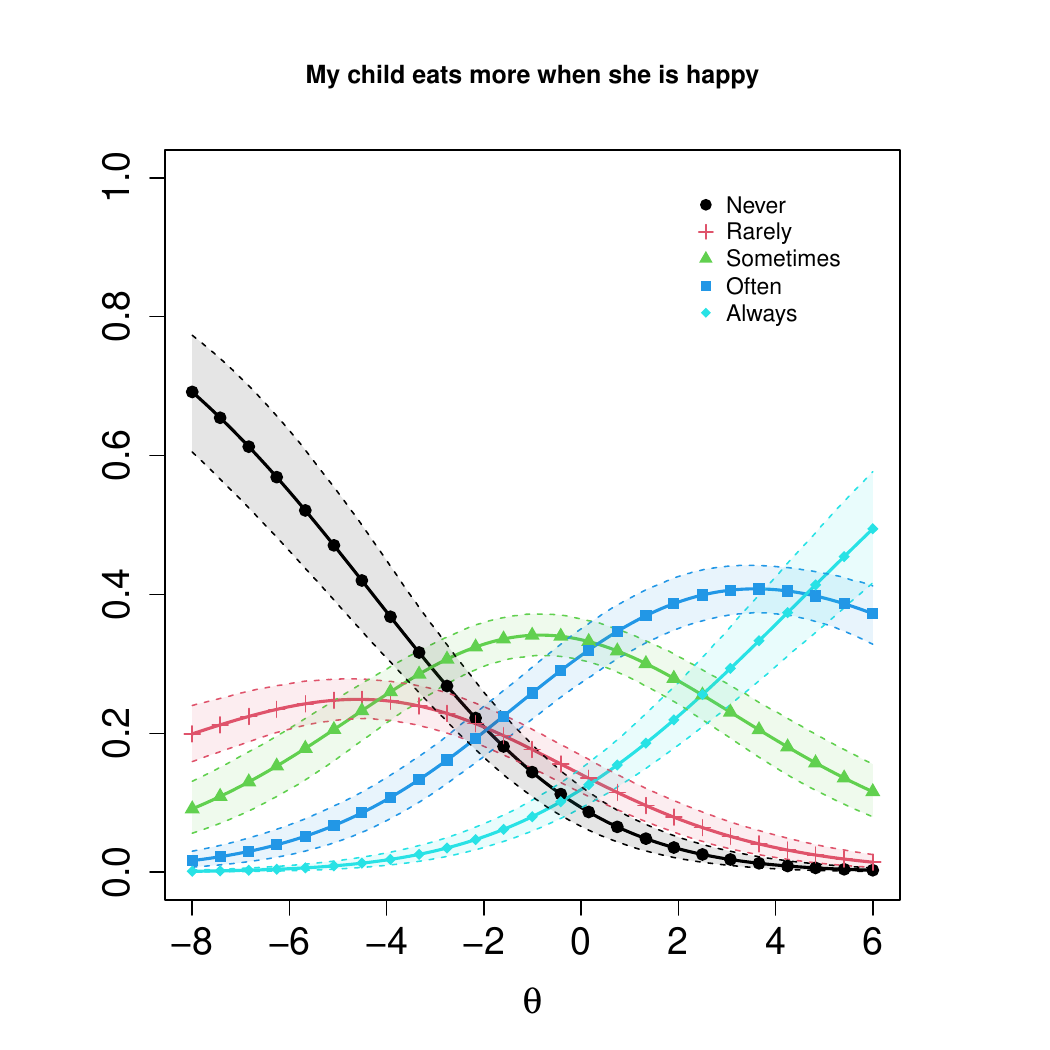}}
\caption{GUSTO data. Posterior item characteristic curves of the five least discriminatory questions, based on the posterior distribution of the corresponding discriminatory parameters $\alpha_j$.}
\label{SMfig:ICC_btm5}
\end{figure}

 \clearpage
	\section{MCMC Algorithm}\label{SMsec:MCMC_Algorithm}
	
We perform the following MCMC updates. Updates for parameters whose full-conditionals are not known in closed form are performed via adaptive Metropolis-Hastings steps \citep{haario2001adaptive}.

\begin{enumerate}
    \item \textbf{Missing observations} in the dataset are sampled from the corresponding full-conditional distribution, which coincides with the sampling model likelihood.
    
    \begin{itemize}
        \item Z-BMI trajectories. Let $I_{miss} = \left(i_1, \dots, i_{n_{miss}}\right)$ be the set of indices corresponding to the missing observations for the $i$-th subject. Note that $I_{miss} = \emptyset$ and $I_{miss} = \left(1, \dots, T_Z\right)$ are admissible index sets, the latter corresponding to the case where all time points are observed. 
        
        Sample $\bm Z_{I_{miss}}$ from the following full-conditional:
        
        $\bm Z_{I_{miss}} \mid \bm Z_{\setminus I_{miss}}, \bm b_i, \bm B_Z, \bm \gamma^Z, \bm X^Z_i, \sigma^2_Z \sim \text{N}_{n_{miss}}\left(\bm \mu^c_Z, \bm \Sigma^c_Z\right)$

        where $\bm Z_{\setminus I_{miss}}$ is the observed part of the vector $\bm Z_i$, while $\bm \mu^c_Z$ and $\bm \Sigma^c_Z$ are the conditional mean vector and covariance covariance matrix obtained by computing the conditional law of the multivariate Gaussian distribution from the likelihood given in Eq.~(2).
        
        \item CEBQ responses. Sample the missing $Y^t_{ij}$ from the following full-conditional probabilities:
        
        $\mathbb{P}\left(Y^t_{ij} = h \mid \theta^{p_j}_{it}, \alpha_j, \bm \beta_j, \bm \gamma^Y_j, \bm X^Y_i\right) =  \!\begin{aligned}[t]
        &\frac{\exp\left(\alpha_{j} \left(h\theta^{p_j}_{it} - \sum_{l = 0}^{h-1} \beta_{jl}\right) + h \bm \gamma^Y_j \bm X^Y_i \right)}{\sum_{h = 0}^{m - 1}\exp\left(\alpha_j \left(h\theta^{p_j}_{it} -  \sum_{l = 0}^{h-1}\beta_{jl}\right) + h \bm \gamma^Y_j \bm X^Y_i \right)}\\
	&h \in \{0, \dots, m-1\}
\end{aligned}$
where $(t,i,j)$ are the indices of the missing response to question $j$ by subject $i$ at time $t$.

    \end{itemize}

    \item \textbf{Longitudinal Z-BMI}

\begin{itemize}
    \item Update $\bm \gamma^Z$ from its full-conditional:

    $$
    \bm \gamma^Z \mid \bm b_i, \bm B_Z, \bm X^Z_i, \sigma^2_Z \sim \text{N}_{q_Z}\left(\bm m_{\bm \gamma^Z}, \bm S_{\bm \gamma^Z} \right)
    $$
with $\bm S_{\bm \gamma^Z} = \frac{\sigma^2_Z}{T_Z} \left(\sum_{i=1}^N\bm X^Z_i (\bm X^Z_i)^T\right)^{-1}$ and $\bm m_{\bm \gamma^Z} = \bm S^{-1}_{\bm \gamma^Z} \sum_{i=1}^N \bm X^Z_i \left(\bm Z_i - \bm b_i \bm B_Z\right)$.

    \item Update $\sigma^2_Z$ from its full-conditional:

    $$
    \sigma^2_Z \sim \text{IG}\left(a_{\sigma^2_Z} + N T_Z / 2, b_{\sigma^2_Z} + \frac{1}{2}\sum_{i=1}^N\sum_{t = 1}^{T_Z}\left(Z_{it} - \bm b_i \left[\bm B_Z\right]_t - \bm \gamma^Z \bm X^Z_i \right)^2 \right)
    $$
    with $a_{\sigma^2_Z} = 3$ and $b_{\sigma^2_Z} = 2$.

\end{itemize}

    \item \textbf{IRT parameters}

Let $p\left(Y^t_{ij}\right) = \mathbb{P}\left(Y^t_{ij} \mid \theta^{p_j}_{it}, \alpha_j, \bm \beta_j, \bm \gamma^Y_j, \bm X^Y_i\right)$, reported earlier. Let also $\bm Y_j = \left(Y^t_{ij}\right)_{i = 1, \dots, N \mbox{ and } t = 1, \dots, T_Y}$.

\begin{itemize}
    \item Update $\theta^{p_j}_{it}$ from the full-conditional distribution proportional to:
    $$
    \log p\left(\theta^{p_j}_{it} \mid Y^t_{ij} = h, \alpha_j, \bm \beta_j, \bm \gamma^Y_j, \bm X^Y_i\right) \propto \log p\left(Y^t_{ij}\right) - \frac{1}{2} \left(\theta^{p_j}_{it} - \theta^{p_j,0}_{it}\right)^2
    $$
    
    \item Update $\alpha_j$ from the full-conditional distribution proportional to:

    $$
    \log p\left(\alpha_j \mid \bm Y_j,\theta^{p_j}_{it}, \bm \beta_j, \bm \gamma^Y_j, \bm X^Y_i\right) \propto \sum_{i=1}^N \sum_{t=1}^{T_Y}\log p\left(Y^t_{ij}\right) - \frac{1}{2} \left(\log\left(\alpha_j\right) - \mu_{s_j}\right)^2 - \log\left(\alpha_j\right)
    $$
    
    \item Update $\bm \beta_j$ from the full-conditional distribution proportional to:

    $$
    \log p\left(\bm \beta_j \mid \bm Y_j, \alpha_j, \bm \theta^{p_j}, \bm \gamma^Y_j, \bm X^Y_i\right) \propto \sum_{i=1}^N \sum_{t=1}^{T_Y} \log p\left(Y^t_{ij}\right) - \frac{1}{2} \bm \beta^T_j \bm \beta_j
    $$
    
    \item Update $\mu_{s_j}$ from the full-conditional distribution:

$$
\mu_s \mid \bm \alpha \sim \text{N}\left(\sum_{j : s_j = s}\log\left(\alpha_j\right)/(n_s + 1), 1/(n_s + 1)\right)
$$
with $n_s = \# \{j : s_j = s\}$ and $s = 1, \dots, n_s$.

    \item Update $\bm \gamma^Y$ from the full-conditional distribution proportional to:

    $$
    \log p\left(\bm \gamma^Y_j \mid \bm Y_j, \bm \theta^{p_j}, \alpha_j, \bm \beta_j, \bm X^Y\right) \propto \sum_{i=1}^N \sum_{t = 1}^{T_Y} \log p\left(Y^t_{ij}\right) - \frac{1}{2} (\bm \gamma^Y_j)^T \bm \gamma^Y_j
    $$

\end{itemize}

    \item \textbf{BNP}

    We employ the sampling scheme by \cite{FavTeh13}, which involves sampling of the auxiliary variable $u$ in an augmented scheme.

\begin{itemize}
        \item Update the allocation variables $\bm c = (c_1, \dots, c_N)$ indicating the cluster assignment of the subjects, such that $i$ belongs to cluster $C_j$ iff $c_i = j$. In order to perform this step, we resort to a P\'{o}lya urn algorithm, conditionally on the auxiliary variable $u$. Indicate with the suffix -$i$ arrays of variables or data where the $i$-th element has been removed, and let the suffix $^{\star}$ indicate the cluster-sepcific values of $\psi^{\star}_j = \left(\bm b^{\star}_j, \bm \theta^{1,0,\star}_j, \dots, \bm \theta^{n_p,0,\star}_j\right)$. Then:
	\begin{align*}	
	&\mathbb{P}(c_i = j|\bm c_{-i}, \bm Z^{-i}, \bm Y^{-i}, \bm X^{-i}, \bm \vartheta) \propto \\
	&\left\{
	\begin{array}{ll} (n^{-i}_j - \sigma)  \text{N}_{T_Z}\left(\bm b^{\star}_j \bm B_Z + \bm \gamma^Z \bm X^Z_i \bm 1_{T_Z}, \mathbb{I}_{T_Z} \sigma^2_Z\right) \prod_{p = 1}^{n_p}\prod_{t = 1}^{T^Y}\text{N}(\theta^p_{it} | \theta^{p,0,\star}_{tj}, 1), & j = 1, \dots, K^{-i}_N \\
	\kappa (1 + u)^{\sigma} \int \text{N}_{T_Z}\left(\bm b \bm B_Z + \bm \gamma^Z \bm X^Z_i \bm 1_{T_Z}, \mathbb{I}_{T_Z} \sigma^2_Z\right) \prod_{p = 1}^{n_p}\prod_{t = 1}^{T^Y}\text{N}(\theta^p_{it} | \theta^{p,0}_t, 1) P_0(d\bm \psi), & j = K^{-i}_N + 1
	\end{array} 
	\right.
	\end{align*}
	with $n^{-i}_j$ representing the size of cluster $j$ after removing the $i$-th element. The expression of $P_0$ is reported in equation (4) in the main text. Due to the non-conjugate setting, we resort to an efficient version Neal's Algorithm 8 \citep{Neal00, FavTeh13}.

        \item Update unique values $\psi^{\star}_j = \left(\bm b^{\star}_j, \bm \theta^{1,0,\star}_j, \dots, \bm \theta^{n_p,0,\star}_j\right)$
$$
\bm b^{\star}_j \mid \bm Z_i, \bm B_Z, \bm \gamma^Z, \bm X^Z_i, \sigma^2_Z \sim \text{N}_d\left(\bm m_{\bm b^{\star}_j}, \bm S_{\bm b^{\star}_j} \right)
    $$
with $\bm S_{\bm b^{\star}_j} = \frac{\sigma^2_Z}{n_j}\left(\mathbb{I}_{d} + \bm B_Z^T \bm B_Z\right)^{-1}$ and $\bm m_{\bm b^{\star}_j} = \bm S^{-1}_{\bm b^{\star}_j} \sum_{i=1}^N \bm BX^Z \left(\bm Z_i - \bm \gamma^Z \bm X^Z_i\right) / \sigma^2_Z$, and

$$
\bm \theta^{p,0,\star}_j \mid \sim \text{N}_{T_Y}\left(\bm \sum_{i : c_i = j} \bm \theta^p_i/(n_j + 1), \mathbb{I}_{T_Y}/(n_j + 1)\right)
$$

        \item Update $u$ from the full-conditional proportional to:

        $$
         (N-1) \log(u) - \kappa/\sigma (1 + u)^{\sigma} - (N - K_N \sigma) \log(1 + u)
        $$
where $K_N$ is the number of clusters and $\kappa$ and $\sigma$ are the hyperparameters of the NGG process. 
    \end{itemize}

\end{enumerate}

	\clearpage
	\bibliographystyle{plainnat}
	\bibliography{Biblio}